\documentclass[%
 reprint,
 amsmath,amssymb,
 aps,
prc,
]{revtex4-2}

\usepackage{graphicx}
\usepackage{dcolumn}
\usepackage{bm}
\usepackage{hyperref}
\hypersetup{
    colorlinks=true,
    citecolor=blue,
    linkcolor=blue,
    filecolor=magenta,      
    urlcolor=blue,
       }

\begin{document}

\preprint{APS/123-QED}

\title{Analysis of the fusion mechanism in synthesis superheavy element 119 via $^{54}$Cr+$^{243}$Am reaction}

\author{B.M. Kayumov$^{1,4}$}
\author{O.K. Ganiev$^{1,2,5}$}
\author{A.K. Nasirov$^{1,3}$}
 \email{nasirov@jinr.ru} 
\author{G.A. Yuldasheva$^{1}$}
\affiliation{
$^1$Institute of Nuclear Physics, Uzbekistan Academy of Science, 100214 Tashkent, Uzbekistan \\
$^2$College of Engineering, Akfa University, 100095 Tashkent, Uzbekistan\\
$^3$Joint Institute for Nuclear Research, 141980 Dubna, Russia \\
$^4$New Uzbekistan University, 100007 Tashkent, Uzbekistan\\
$^5$Faculty of Physics, National University of Uzbekistan, 100174 Tashkent, Uzbekistan
}

\date{\today}

\begin{abstract}
The combined dinuclear system (DNS) and statistical model implanted in KEWPIE2 have been used to study the prospects for the synthesis of a 
superheavy element (SHE) with $Z=119$ in the $^{54}$Cr+$^{243}$Am fusion reaction. The method of calculation has been verified by 
description of the evaporation residue cross sections measured for the $^{48}$Ca+$^{243}$Am reaction. The calculated results of the 
partial and total cross sections for the complete fusion, quasifission, fast fission and evaporation residues formation for both reactions are 
discussed.
\end{abstract}

\maketitle


\section{\label{sec:level1}Introduction}

In modern nuclear physics, the fusion of massive nuclei is one of the main research topics due to the great motivation in the synthesis of  
new 	superheavy elements (SHEs) \cite{Hofmann2000,Armbruster2000,Morita2004,Oganessian2006,Oganessian2007,		
Oganessian2005,Oganessian2012,Oganessian2013,Munzenberg2015}. Remarkable recent progress in the synthesis of new SHEs has 
been achieved using of the $^{48}$Ca projectile with the transactinides \cite{Oganessian2006,Oganessian2007,Oganessian2005,Oganessian2012,Oganessian2013}.  
Various theoretical and experimental studies have been carried out to study the nuclear fusion process of the formation of SHE aiming to two main aspects:
 the first reason is to find out and analyze the mechanism of synthesis of SHE,
  the second is to search for the best projectile-target combination 	
and a suitable incident energy  for obtaining new SHEs and their isotopes. In particular, the evaporation residue (ER) cross section  
strongly depends on the incident energy and the projectile-target combination. Therefore, the study of such dependences is interesting 
and especially useful when trying to synthesize new SHEs with $Z>118$. 
This is important due to the fact that the ER cross section of fusion 
reactions with these nuclei becomes very small, and the excitation functions are very narrow. This circumstance causes  extremely 
difficulties in synthesis of the 	heaviest superheavy elements in experiments. 

In order to determine the best conditions for formation of SHE, several experiments were carried out to study the influence of the 
entrance channel on the ER cross section \cite{Dvorak2009,Graeger2010,Nishio2012}. The isospin effect of the target nucleus has been 
recently investigated on the ER cross section \cite{Brewer2018}. In some laboratories, the experiments aimed to synthesize SHEs with Z 
= 119 and 120 were performed  via hot fusion reactions \cite{Oganessian2009,Hofmann2011}. To estimate the fusion probability correctly 
it is necessary to distinguish pure fusion-fission products from the quasifission products 
\cite{Toke1985,duRietz2013,Itkis2007,Itkis2011,Kozulin2014,Giardina2018}. Since the experimental characteristics of the quasifission and 
fusion-fission processes \cite{Banerjee2019} may be similar. Therefore, it is important a better understanding the dependence of the 
complete fusion process on the entrance channel to distinguish mechanisms creating fusion-fission and quasifission fragments which can 
have overlap in the observed mass distributions.
	
In general, many theoretical models have been developed and applied to interpret experimental data by estimating the evaporation 
residue cross section. On one hand, the synthesis mechanism of SHE needs to be elucidated 
\cite{Adamian1997,Shen2008,Zagrebaev2008,Wang2008,Liu2009,Nasirov2009,Smolanczuk2010, Wang2011,Siwek-Wilczynska2012,Wang2014,Bao2015a,Wu2018,Huang2011}. Various approaches are devoted to calculating the fusion probability of the 
colliding nuclei, as well as analyzing the distribution of quasifission fragments 
\cite{Nasirov2009,Antonenko1993,Antonenko1995,Nasirov2007}. On the other hand, in order to produce the new SHEs, or their isotopes 
far from the island of stability of the chemical elements, it is necessary to estimate the corresponding incident energy and the best 
projectile-target combination.
	
The extended nuclear landscape allows us to investigate the nuclear structure of SHE and the nuclear reaction mechanism. To search for 
the optimal condition of synthesis, the influence of the entrance channel \cite{Nasirov2009,Nasirov2005,Nasirov2011,Hong2016} and the 
isospin of heavy colliding nuclei \cite{Mandaglio2012,Bao2015b} on the evaporation 	residual cross section have been studied systematically 
in many works. In particular, in some works	\cite{Nasirov2009,Shen2008,Wang2012,Liu2013,Zhu2014,Bao2016,Li2018,XingLv2021} predictions 
were made of a possible method for the synthesis of new SHEs with $Z$ = 119 and 120.
	
On the whole, the final cross section for the formation of SHE depends on two factors: 1) the fusion probability in the entrance channel; 2) 
competition between the fission process and the evaporation of neutrons in the output channel. In case of the massive nuclei interaction, 
the capture of the projectile-nucleus by the target-nucleus 	does not lead directly to their complete fusion as in case of the 	collision of the 
light nuclei. Alternative channel to complete fusion is quasifission process when the DNS formed at capture breaks down after 
multinucleon process due to presence of the intrinsic fusion barrier against transfer all nucleons of the lighter fragment to the heavy one. 
The maximum of the mass-charge distribution of the quasifission products is concentrated near yield of the magic nuclei having the proton 
and/or neutron magic numbers. Experiments with massive nuclei show that quasifission is dominant channel in comparison with complete 
fusion, particularly in cold fusion reactions \cite{Kozulin2014,Nasirov2014}.
 Therefore, the fusion probability becomes very sensitive to the entrance 
channel. The competition between the fission process and the evaporation of neutrons at each stage of the decay process will significantly 
reduce the survival probability of the final SHE. Therefore, accurate calculations and predictions of these cross sections are very important 
since experimenters can choose the optimal combinations of projectile and target, as well as the favorable beam energy at which the 
synthesis of SHE is possible. In particular, in order to study the formation of evaporation residues, it is important to analyze the role of  
the entrance channel characteristics, such as incident energy and orbital angular momentum which are responsible at formation the 
angular momentum distribution of the excited compound nucleus. Thus, this research could be useful in planning future experiments.

The main goal of this paper is to examine the possibility of the formation of the 119th superheavy element as a result of the $^{54}$Cr+
$^{243}$Am reaction within the framework of the developed DNS model. In particular, we have calculated the capture, fusion, and ER 
cross sections for the $^{48}$Ca+$^{243}$Am and 	$^{54}$Cr+$^{243}$Am reaction based on the developed DNS model.	We have 
demonstrated the role of the orbital angular momentum in the formation of a compound nucleus.

\section{\label{sec:level2}Theoretical description for the formation of SHN}

\begin{figure*}[htbp]
  \centering
  \includegraphics[width=1\textwidth]{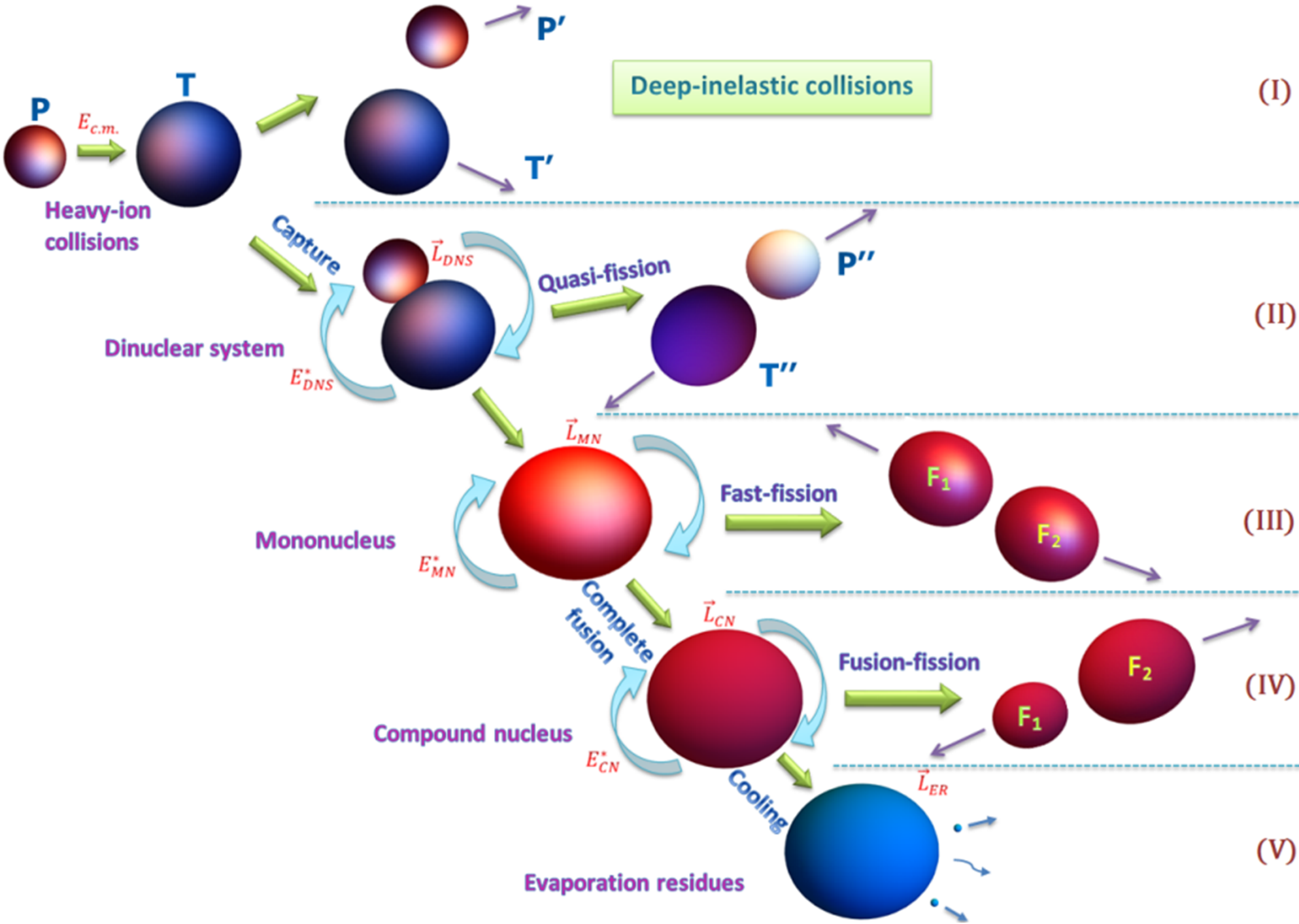}
\caption{(Color online) Schematic diagram of the reaction mechanisms 
causing hindrance at formation of the superheavy nuclei in 
heavy-ion collisions. The multinucleon transfer reactions and fission 
 (I--IV) leading to the formation of binary or fission-like 
fragments, which compete in the way to the ER formation (superheavy nuclei).
I. deep-inelastic collision (not full momentum transfer) of the initial projectile
 P and target T nuclei;
II. Quasifission (full momentum transfer) products P'' and T''; 
III. Fast fission (fission of non-equilibrated mononucleus) products F$_1$ and 
F$_2$;
IV. Fusion-fission (fission of compound nucleus) products F$_1$ and F$_2$.} 
\label{sketch}
\end{figure*}

The formation of SHE is a result of the complicated dynamical process with the multinucleon transfer between interacting nuclei. Therefore, 
it is very important to better understand the different steps of the reaction mechanism of SHE formation. Theoretically, from our point of 
view, the SHE synthesis process can be divided into several reaction stages. A schematic diagram of the stages preceding the formation 
of evaporation residues is shown in Fig. \ref{sketch}. The first stage is a competition between deep-inelastic collision (incomplete 
momentum transfer) and capture process (full momentum transfer). At this reaction stage, the colliding nuclei overcome the Coulomb 
barrier and come into close contact with the overlapping surfaces of the nuclei. As a function of the initial collision energy and impact 
parameter one of above mentioned events takes place. If after dissipation of the sufficient part of the relative kinetic energy the system 
is able overcome the Coulomb barrier from the inside to outside the deep-inelastic collision (I)  occurs (see Fig.\ref{sketch}) and if the 
residual part of the	kinetic energy is not enough to overcome the Coulomb barrier the system is trapped into potential well of the nucleus-
nucleus interaction (see Fig. 2 in Ref.~\cite{Nasirov2005}. The last case 	is 
 called capture and full momentum transfer takes place (II). The difference between capture and 
deep-inelastic collisions depends on whether the path of relative motion has been trapped into the potential well or not. Consequently, the 
lifetime of DNS formed at capture is longer in comparison the one of DNS formed in the deep-inelastic collisions. This competition strongly 
depends on the collision energy and the impact parameter (angular momentum of relative motion). This fact allows to separate the 
products of these two different reaction mechanisms.
In the second stage of the reaction, the evolution of DNS can end with the CN formation with excitation energy $E_{CN}^*$ in an 
equilibrium state, or it can arrive at two fragments without reaching the state of CN. The last process is called quasifission (II) (see Fig. 
\ref{sketch}). The quasifission occurs when the DNS prefers to break down into two fragments rather than becoming a fully equilibrated 
CN. The quasifission is the main hindrance to the CN formation during the evolution of DNS in reactions with massive nuclei, since this 
process is dominant relative to the complete fusion. This channel of evolution of DNS due to multinucleon transfer is shown as channel II. 
Quasifission products can have characteristics similar to those of nuclear fission. For example, the mass distributions of the fusion-fission and quasifission
products may overlapped parts and the total kinetic energy of quasifission 
fragments is close to the total kinetic energy of fission fragments. 
The third stage is the process of formation of a rotationally excited 
 mononucleus, which occurs due to the transfer of all nucleon from the light fragment to heavy nucleus. Rotation of a non-equilibrium 
mononucleus causes decreasing  its fission barrier. 
Consequently, at a certain value of the orbital angular momentum, the fission 
barrier disappears completely. In this case, the non-equilibrium mononucleus becomes unstable and undergoes fission from some excited 
state. The splitting of a mononucleus before reaching its equilibrium state is called fast-fission (III), as a result of which binary mass-
symmetric fragments are formed. This process leads to the formation of binary products (see Fig. \ref{sketch}), which causing 
hindrance for the heated and rotating mononucleus to be de-excited and transformed into CN. So, fast fission depends on the splitting of 
the mononucleus before reaching equilibrated CN, if $L_{CN}>L_f$ \cite{Sierk1986}, where $L_f$ is a value at which fission barrier of CN 
disappears at the given excitation 	energy. It is well known that the increase of the CN excitation energy also decreases of the fission 
barrier appeared due to shell effects in massive nuclei.  This phenomenon 
is related with the washing out the shell effects in the heated nuclei.

Recall that the quasifission process can occur for all $L_{DNS}$ values of the DNS. This is one of the main differences between fast-fission 
and quasifission. The fourth stage is that the binuclear system evolves from a tangent configuration to the CN formation, which can be 
estimated by the probability of merging. In the last stage, the excited CN undergoes to the last competition between fission into two 
fragments (IV) and cooling by the emission of neutrons, and the result of this competition is estimated by the survival probability. It 
should be stressed that a very small evaporation residue cross section is obtained for the SHN production (V).

\section{\label{sec:level3}Outline of the approach}

The study of the main processes taking place in heavy-ion collisions near 	the Coulomb barrier energies is based on calculations of the 
incoming path 	of the projectile nucleus and finding the capture probability for the interaction with different orientation angles of the axial 
symmetry axis of a deformed nuclei and at the given value of the collision impact 
 parameter  
\cite{Nasirov2005,Fazio2008,Nasirov2013,Kim2015}. Also, the surface vibrations of the nuclei, which are spherical in the ground state  are 
taken into consideration. The final results are averaged over all orientation angles ($\alpha_1$ and $\alpha_2$) of the axial symmetry axis 
of deformed nuclei relative to the beam direction or vibrational states of the spherical nuclei. These procedures are presented in Subsection 
\ref{subsec:level6} of this work.

\subsection{\label{sec:level4}Capture}

The capture process in heavy-ion collisions (i.e., capture of a projectile nucleus
 by a target nucleus) is closely related to the  
peculiarities  of interacting nuclei and the entrance channel effects. This is due to the fact that the nucleus-nucleus potential depends  on 
the shape and orientation angles of the symmetry axis of the interacting nuclei and	 plays an important role in determining the DNS fate. 
In generally, a DNS consists of two basic degrees of freedom that lead to capture: (i) the intrinsic degree of freedom due to the exchange 
of nucleons between the projectile nucleus and the target, and (ii) the collective degree of freedom from relative motion between the two 
nuclei leads to capture. In this context,  the capture cross section is one of primary interest to experimentalists and theorists studying the 
reaction mechanism in heavy-ion collisions since only full momentum transfer to the intrinsic and collective degrees of freedom from the 
relative motion is the beginning of the way of the compound nucleus formation \cite{Nasirov2013,Kim2015,Nasirov2016}. The capture 
cross section is an observable quantity and, at the same time, one of the important components in the synthesis of superheavy nuclei. In 
this regard, in order to successfully predict a possible way to synthesize the new SHEs, it is very important to carefully examine the 
capture process associated with an accurate description of the capture cross section measurements.

Typically, capture requires two conditions to be met:
\begin{enumerate}
		\item The initial energy of the projectile nucleus in the center-of-mass system $E_{\rm c.m.}$ must be sufficient to reach the potential well by 
overcoming or tunneling through the barrier at a relative distance $R$ in the entrance channel;
		\item Some of the relative kinetic energy must be dissipated for the dinuclear system to induce trapping in the potential well.
	\end{enumerate}
	
This condition depends on the shape and orientation angles ($\alpha_1$ and $\alpha_2$) of the symmetry axis of the interacting nuclei 
and orbital angular momentum ($L=\ell \hbar$) which determine the	 size of the potential well, and the intensity of the friction forces that 
cause the kinetic energy of the relative motions to dissipate into the internal energy of two interacting nuclei. In our approach, the height 
of the inner barrier of the well of the nucleus-nucleus potential is called the quasifission barrier which is a hindrance against decay of 
dinuclear system. An alternative process of quasifission in the evolution of the DNS is the complete fusion of its constituent fragments. 
Thus, the capture cross section for a given relative energy in the center of mass system ($E_{\rm c.m.}$) and angular momentum ($\ell$) is the sum of the cross sections for complete fusion, quasifission and fast fission:
\begin{eqnarray}\label{cap}
\sigma_{cap}(E_{\rm c.m.},\ell;\alpha_1,\alpha_2) &=& 
\sigma_{fus}(E_{\rm c.m.},\ell;\alpha_1,\alpha_2)\nonumber\\
&+&\sigma_{qfis}(E_{\rm c.m.},\ell;\alpha_1,\alpha_2)\nonumber\\
&+&\sigma_{ffis}(E_{\rm c.m.},\ell;\alpha_1,\alpha_2).
\end{eqnarray}
The appearance of the term $\sigma_{ffis}(E_{\rm c.m.},\ell;\alpha_1,\alpha_2)$ presenting a contribution of the fast fission process in nucleus-nucleus collisions is discussed in Section \ref{Discussion}.

The capture cross section is determined by the number of partial waves which lead the path of the total energy of colliding nuclei to be trapped in the well of the nucleus-nucleus potential after dissipation of a sufficient part of the initial kinetic energy \cite{Kim2015}. 
The size of the potential well decreases with increasing orbital angular momentum $\ell$.

The capture cross section is calculated by the following formula:
\begin{eqnarray}\label{capture}
\sigma_{cap}(E_{\rm c.m.},\ell;\alpha_1,\alpha_2)&=&\frac{\lambda^2}{4\pi}\sum_{\ell=0}^{\ell_d}(2\ell+1)\nonumber\\
&\times&\mathcal{P}^{\ell}_{cap}(E_{\rm c.m.},\ell;\alpha_1,\alpha_2),
\end{eqnarray}
where $\lambda$ is the de Broglie wavelength of the entrance channel and $\mathcal{P}^{\ell}_{cap}(E_{\rm c.m.},\ell;\alpha_1,\alpha_2)$ 
is the capture probability which depends on the collision dynamics:
\begin{widetext}
\begin{eqnarray}
\mathcal{P}^{(\ell)}_{cap}(E_{\rm c.m.},\ell,\{\beta_i\}) = \left\{
\begin{array}{lll}
1,\,\, \mbox{if}\,\, \ell_m < \ell < \ell_d \,\, \mbox{and} \,\, E_{\rm c.m.} > V_{B}, \\
0,\,\, \mbox{if} \,\,\ell < \ell_m \,\, \mbox{or}\,\,\ell>\ell_d \,\, \mbox{and} \,\, E_{\rm c.m.}>V_{B}\, \\
\mathcal{P}^{(\ell)}_{tun}(E_{\rm c.m.},\ell,\{\beta_i\}),\,\, \mbox{for all}\,\, \ell \mbox{ if } E_{\rm c.m.}\leq V_{B}\, \\
\end{array} \right.
\label{CapClass}
\end{eqnarray}
\end{widetext}
where $V_B$ is the barrier of the nucleus-nucleus potential in the entrance channel; $\beta_i$ are parameters of the nuclear shape 
deformation; $\ell_m$ and $\ell_d$ are the minimum and maximum values of the orbital angular momentum $\ell$ leading to capture at 
the given collision energy. This means that the values of $\ell$ leading to capture can form a ``window''	determined by the size of the 
potential well of the nuclear-nuclear potential. The fact is that the coefficient of friction is not so hight enough to trap the projectile in a 
potential well. Thus, the boundary values $\ell_{min}$ and $\ell_d$ of the partial waves leading to capture depend on the dynamics of 
collision and they are determined by solving the equations of motion for the relative distance $R$ and orbital angular momentum $\ell$ 
\cite{Nasirov2005,Fazio2003}. $\mathcal{P}^{(\ell)}_{tun}$ is probability of the barrier penetrability which is calculated by the improved 
WKB formula by Kemble et al \cite{Kemble1935}:
\begin{eqnarray}\label{penetration}
\mathcal{P}^{(\ell)}_{tun}(E_{\rm c.m.},\{\beta_i\})=\frac{1}
{1+\exp\left[2K(E_{\rm c.m.},\ell,\{\beta_i\})\right]},
\end{eqnarray}
where
\begin{eqnarray}
K(E_{\rm c.m.},\ell,\{\beta_i\})=\int\limits_{R_{in}}^{R_{out}}\sqrt{\frac{2\mu}{\hbar^2}(V(R,\ell,\{\beta_i\})-E_{\rm c.m.})}\,dR.\nonumber\\
\end{eqnarray}
$R_{in}$ and $R_{out}$ are inner and outer turning points which were estimated by $V(R)=E_{\rm c.m.}$.

\subsection{\label{sec:level5}Potential energy surface. The driving potential}

\begin{figure}[htb]
	\centering
	\includegraphics[width=0.55\textwidth]{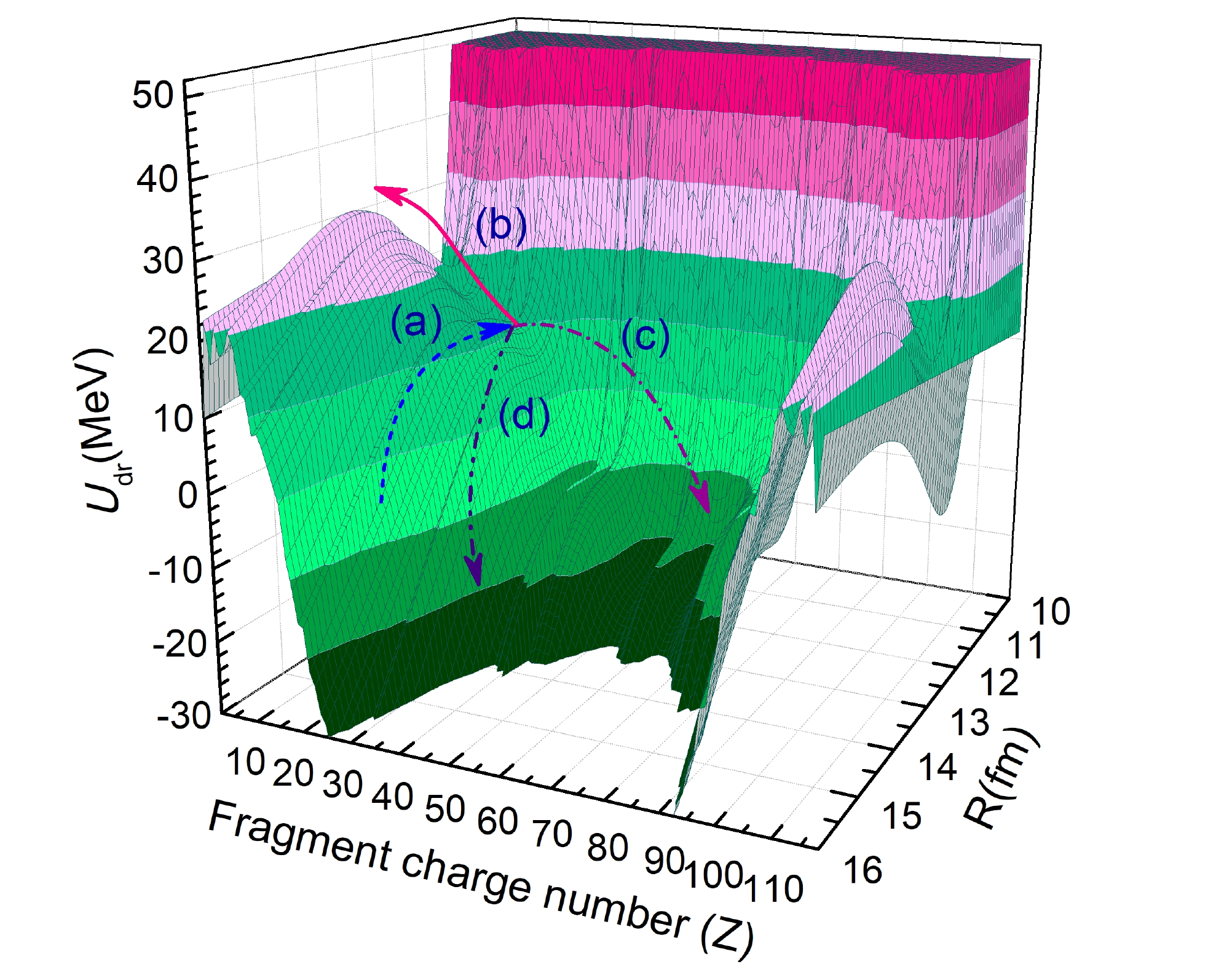}
	\caption{\label{PES_Cr+Am}(Color online) Potential energy surface calculated as a function of the relative distance between the interacting 
		nuclei and fragment charge number  for the $^{54}$Cr+$^{243}$Am reaction 
		leading to the CN $^{297}119$. Arrow (a) shows the  capture path in 
		the entrance channel; arrow (b) shows the direction of the complete fusion by multinucleon transfer from the light nucleus to the heavy 
		one; (c) and (d) arrows show the directions of decay of the DNS into mass symmetric and asymmetric quasifission channels, 
		respectively.}       
\end{figure}

The potential energy of a heavy nuclear configuration plays a crucial role in understanding the evolution of collisions of nuclei. Generally, the 
potential energy surface (PES) is very important in the DNS model, which provides information about the optimal projectile-target 
combination, the optimal excitation energy, and influences the fusion probability significantly. The PES is calculated as a function of the 
mass and charge numbers of its fragments, orbital angular momentum ($\ell$) of collision and relative distance $R$ between their mass 
centres (see Fig.~\ref{PES_Cr+Am}):  
\begin{eqnarray}\label{drivpot}
U(Z_i,A_i,\ell,R,\beta_i,\alpha_i)&=&Q_{gg}
+V_N(Z_i,A_i,\ell,R,\beta_i,\alpha_i)\nonumber\\
&-&V_{rot}^{CN}(\ell) \quad (i=1,2),
\end{eqnarray}
where $\beta_i (i = 1,2)$ represent quadrupole deformations of the colliding nuclei, $Q_{gg}=B_1+B_2-B_{CN}$ is the reaction energy 
balance that is used to determine the excitation energy of CN; $B_1$, $B_2$, and $B_{CN}$ are the binding energies of the interacting 
nuclei and CN, respectively, which are obtained from the nuclear mass tables in Refs. \cite{Audi1995,Moller1995}. 
And also nucleus-
nucleus potential ($V_N$) consists of three parts:
\begin{eqnarray}\label{NNpot}
V_N(Z_i,A_i,\ell,R,\beta_i,\alpha_i)&=& V_{nucl}(Z_i,A_i,\ell,R,\beta_i,\alpha_i)\nonumber\\
&+& V_{Coul}(Z_i,A_i,\ell,R,\beta_i,\alpha_i)\nonumber\\
&+& V_{rot}(Z_i,A_i,\ell,R,\beta_i,\alpha_i),
\end{eqnarray}
where $V_{nucl}$, $V_{Coul}$, and $V_{rot}$ are the nuclear, Coulomb, and rotational potentials, respectively. 
The methods of 
calculation of these three parts of the nucleus-nucleus potential as a function of the orientation angles of the symmetry axis of the colliding 
nuclei are presented in Appendix A of Refs. \cite{Nasirov2005,Fazio2003}.  
If the colliding nuclei are deformed then the parameters ($\beta_i, i=1,2$)  of their deformation and orientation angles ($\alpha_i$) of axial 
symmetry axis are included into consideration. 

Fig.~\ref{PES_Cr+Am} shows the PES calculated at $\ell=0$ for $^{54}$Cr+$^{243}$Am reaction leading to  the CN $^{297}119$. 
The arrow (a) shows the capture path in the entrance channel; the arrow (b) shows the direction of complete fusion by multinucleon 
transfer from the light nucleus to the heavy one; the  arrows (c) and (d) show the directions of decay of the DNS into mass symmetric 
and asymmetric quasifission channels, respectively. Only some of the paths of the DNS evolution along the charge asymmetry axis (the 
solid arrow (b)) and surviving against decay along the relative distance lead to complete fusion, i.e., to the CN formation.
\begin{figure}[htb]
  \centering
  \includegraphics[width=0.48\textwidth]
  {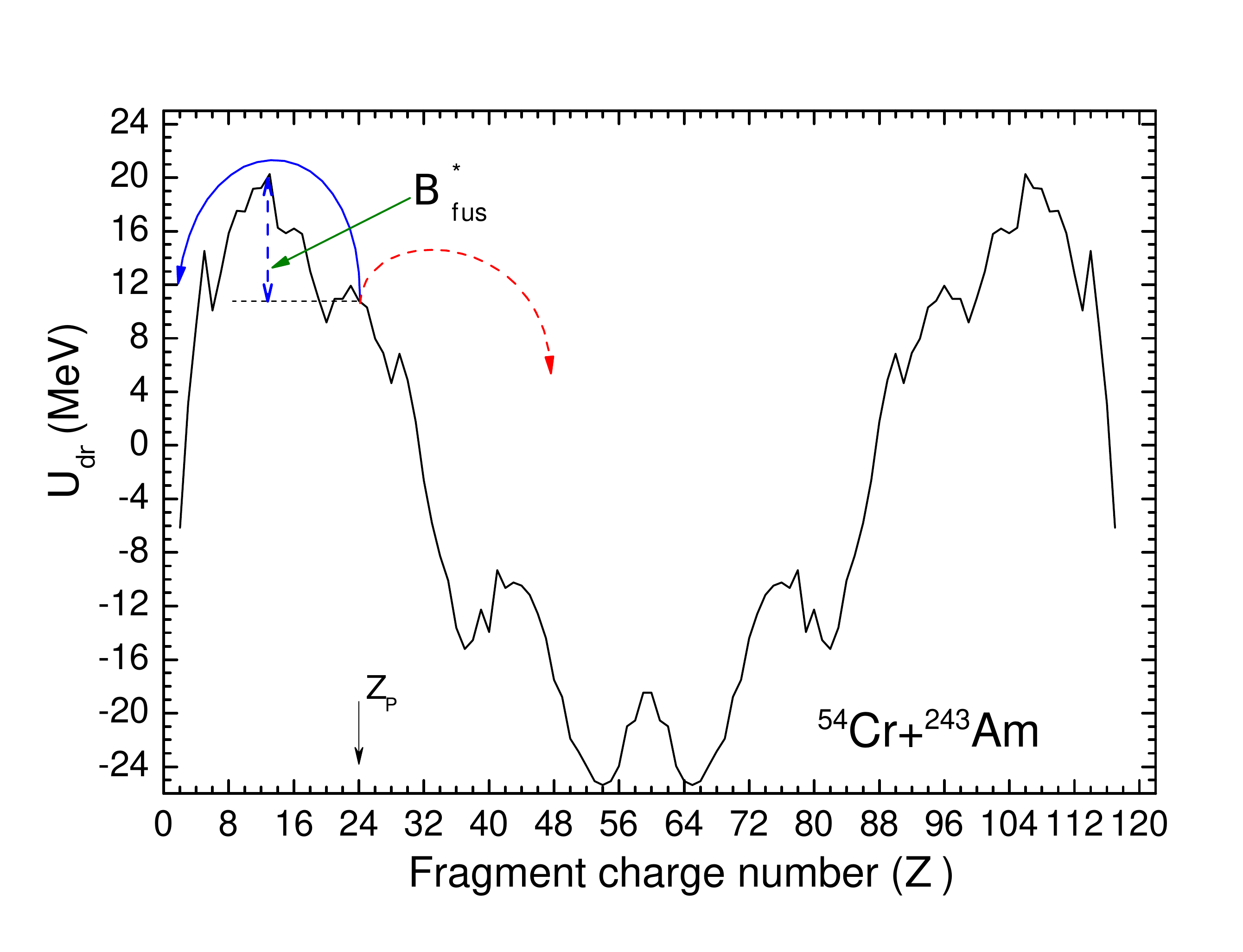}
\caption{(Color online) Calculated driving potential for the DNS 
$^{297}119$ formed in the $^{54}$Cr+$^{243}$Am reaction versus the charge number of its fragment. $B^*_{\rm fus}$ is the intrinsic fusion barrier causing 
a hindrance to complete fusion. The solid curved arrow shows the way to complete fusion while the dashed curved arrow shows evolution of DNS to quasifission channels.}
\label{DrP}
\end{figure}
The intrinsic barrier $B^*_{fus}$ causing a hindrance to complete fusion and quasifission barrier $B_{qf}$ preventing a decay of DNS are 
determined from the landscape of PES. The intrinsic barrier $B^*_{fus}$ is estimated from the peculiarities of the driving potential which is 
formed as a curve connecting the minima of the well of the nucleus-nucleus potential corresponding to the different mass asymmetry 
configuration of dinuclear system formed at capture (see Fig. \ref{DrP}). In Fig. \ref{DrP}, the $B^*_{fus}$ value for the DNS 
configuration $^{54}$Cr+$^{243}$Am  is demonstrated. As a quasifission barrier $B_{qf}$ the depth of the well of the nucleus-nucleus 
potential is used. These quantities are used in calculation of the fusion probability $P_{CN}$ of the interacting DNS fragments.

\subsection{\label{subsec:level6}Fusion}

In calculation of the fusion probability the change of the mass and charge distributions during capture is included, {\it i.e.} the fusion  
probability of the colliding nuclei is taken as a sum of the competition of complete fusion and quasifission at different charge asymmetry  
from the symmetric configuration $Z_{sym}$ of DNS up to configuration corresponding to the maximum value of the driving potential 
$Z_{max}$:
\begin{eqnarray}
P_{CN}(E_{\rm c.m.},\ell,\{\beta_i\})=\sum\limits^{Z_{max}}_{Z_{sym}}D_Z(E^*_Z,\ell)P^{(Z)}_{CN}(E^*_Z,\ell,\{\beta_i\}).\nonumber\\
\label{PCNZ}
\end{eqnarray}
The weight function $D_Z(E^*_Z,\ell)$ is the  charge distributions probability in the 	DNS fragments which is determined by solution of the 
transport master equation \cite{Nasirov2016}. The fusion probability $P^{(Z)}_{CN}(E^*_Z,\ell,\{\beta_i\})$ for the DNS fragments with 
the charge configuration $Z$ rotating with the orbital angular momentum $\ell$ is calculated as the branching ratio $P^{(Z)}_{CN}(E^*_{Z},\ell;\{\alpha_i\})$ of widths related to the overflowing over the quasifission barrier $B_{qf}(Z)$ at a given mass 
asymmetry, over the intrinsic barrier $B_{fus}(Z)$ on mass asymmetry axis to complete fusion \cite{Zhang2010}:
\begin{widetext}
\begin{eqnarray}\label{PcnG}
P^{(Z)}_{\rm CN}(E^*_Z,\ell)\approx{\frac{{\Gamma^{(Z)}_{\rm fus}}(B^{*(Z)}_{\rm fus},E^*_Z,\ell)}{{\Gamma^{(Z)}_{(\rm qf)}(B^{(Z)}_{\rm qf},E^*_Z,\ell)+\Gamma^{(Z)}_{(\rm fus)}(B^{*(Z)}_{\rm fus},E^*_Z,\ell)+\Gamma^{(Z)}_{\rm sym}(B^{(Z)}_{\rm sym},E^*_Z,\ell)}}},
\end{eqnarray}
\end{widetext}
where $\Gamma_{\rm fus}$,  $\Gamma_{\rm qf}$ and  $\Gamma_{\rm sym}$ are corresponding widths determined by the level 
densities on the barriers  $B^{(Z)*}_{\rm fus}$, $B^{*(Z)}_{\rm sym}$  and $B^{(Z)}_{\rm qf}$ are the intrinsic fusion, symmetric 
fission and quasifission barriers; $E^{*}_{Z}$ is the the excitation energy of DNS with the charge asymmetry $Z$ and it is calculated by 
formula 
\begin{eqnarray}
\label{Edns} E^*_Z(A,\ell) &=& E_{\rm c.m.}\nonumber\\
&-& V(Z,A,R_m,(\ell))+\Delta Q_{\rm gg}(Z,A),
\end{eqnarray}
where $V(Z,A,R_m,(\ell))$ is the minimum value of the potential well $V(Z,A,R,\ell)$ at $R_m$;  $\Delta Q_{\rm gg}(Z,A)=B_1+B_2-B_P-
B_T$ is included to take into account the change of the intrinsic energy of DNS due to nucleon transitions during its evolution along mass 
and charge asymmetry axes, where $B_P$, $B_T$, $B_1$ and $B_2$  are binding energies 	of the initial (``P'' and ``T'')  and (``1'' and ``2'') 
interacting fragments at the given time $t$ of interaction. The width $\Gamma_i$  of the decay through the corresponding barriers $B_i$ 
is determined by the level densities on the barriers \cite{Nasirov2005}:
\begin{eqnarray}
\label{Gammai}
\Gamma^{(Z)}_i(B^{*(Z)}_i,E^*_Z,\ell)\approx\frac{\rho_i(E^*_Z-B^{*(Z)}_i,\ell)}{\rho_i(E^*_Z,\ell)}T_Z,
\end{eqnarray}
 where $T(Z) = 3.46 \sqrt{E^*_Z/(A_{P}+A_{T})}$ is the effective temperature of the DNS having the charge asymmetry $Z$.
 As a level density $\rho(E^*_Z,\ell)$ is calculated by the expression~\cite{Antonenko1995}:
\begin{eqnarray}
\rho(E^*_Z,\ell)=\left[\frac{g^2}{g_1 g_2}\right]^{1/2}
\frac{g}{6^{3/4}(2gE^*_Z)^{5/4}}\exp\left[2(aE^*_Z)^{1/2}\right],\nonumber\\
\end{eqnarray}
where $2g=g_1+g_2$ and $\pi^2g_i/6={A_i}/11$ MeV$^{-1}$. 

The partial fusion excitation function is  determined by the partial capture cross sections $\sigma^{\ell}_{cap}$ and   fusion  probabilities $P_{CN}(E_{\rm c.m.},\ell,\{\beta_i\})$ of DNS for the heavy collisions with the given orbital angular momentum $\ell$:
\begin{eqnarray}
\label{parfus}
\sigma_{fus}(E,\ell,\alpha_1,\alpha_2)=\sigma_{cap}(E,\ell;\alpha_1,\alpha_2)P_{CN}(E,\ell;\alpha_1,\alpha_2).\nonumber\\
\end{eqnarray}
Partial fusion cross section is found by averaging over all values of the orientation angles $\alpha_1$ and $\alpha_2$ symmetry axis of 
the deformed projectile and target nuclei:
\begin{eqnarray}\label{parfusal}
\sigma_{fus}(E^*_{CN},\ell)&=&\int\limits_{0}^{\pi/2}\int_{0}^{\pi/2}\sigma_{fus}(E^*_{CN},\ell;\alpha_1,\alpha_2)\nonumber\\
&&\times\sin{\alpha_1}\sin{\alpha_2}d\alpha_1 d\alpha_2.
\end{eqnarray}

To calculate the capture and fusion cross sections in the reactions with the nuclei of the spherical shape their surface vibrations have 
been considered as independent harmonic vibrations and the nuclear radius is taken to be distributed as a Gaussian
distribution~\cite{EsbensenNPA352},
\begin{eqnarray}
g(\beta, \alpha) = \exp
\left[ -\frac{(\sum_{\lambda}\beta_{\lambda} Y_{\lambda 0}^* (\alpha))^2}{2 \sigma_{\beta}^2} \right] (2\pi \sigma_{\beta}^2)^{-1/2},
\end{eqnarray}
where $\alpha$ is the direction of the spherical nucleus. For simplicity, we use $\alpha=0$:

\begin{eqnarray}
\sigma^2_{\beta} = R_0^2 \sum_{\lambda}\frac{2\lambda + 1}{4\pi} \frac{\hbar}{2D_\lambda \omega_\lambda} = \frac{R_0^2}{4\pi} \sum_{\lambda} \beta_\lambda^2,
\end{eqnarray}
where $\omega_{\lambda}$ is the frequency and $D_{\lambda}$ is the mass parameter of a collective mode.

As the amplitudes of the surface vibration we use deformation parameters of first excited 2$^+$ and $3^-$ states of the colliding nuclei. 
The values of the deformation parameters of first  excited 2$^+$ and $3^-$ states  are presented in 
Table \ref{tabdeform} which are taken from Ref(s).~\cite{Raman} ($\beta^+_2$) and \cite{Spear} ($\beta^-_3$).

Partial fusion cross section is found by averaging over values of the vibrational states $\beta_i$  of the spherical  nuclei:
\begin{eqnarray}
\label{parfusav}
\sigma_{fus}(E^*_{CN},\ell)=\int\limits_{-\beta_0}^{+\beta_0}\sigma_{fus}(E^*_{CN},\ell;\beta)
g(\beta)d\beta.
\end{eqnarray}

The  partial fusion cross sections calculated by this way are used in estimations 
of the partial evaporation residue cross sections for 
the given values of the angular momentum distributions of the compound nucleus
 formed with the excitation energy 
\begin{equation}
E^*_{CN}(\ell)=E_{c.m.}+Q_{gg}-V^{\ell}_{CN},  
\end{equation}
where $Q_{gg}=B_P+B_T-B_{CN}$ is a reaction energy balance;  $V^{\ell}_{CN}$  and  $B_{CN}$ are rotational and binding energies 
of the compound nucleus. 
\begin{table}[b]
	\caption{Deformation parameters $\beta_2$ and $\beta_3$ of first excited
		2$^+$ and $3^-$ states of the colliding nuclei used in the calculations in this work.
		\label{tabdeform}}
\begin{ruledtabular}
\begin{tabular}{cccc}
				Nucleus  &  $^{48}$Ca &  $^{54}$Cr & $^{243}$Am \\
				\hline
				$\beta^+_2$~\cite{Raman} & 0.106  & 0.250   & 0.293 \\
				$\beta^-_3$~\cite{Spear}  & 0.101    & 0.250    & 0.083 \\
\end{tabular}
\end{ruledtabular}
\end{table}

\subsection{\label{subsec:3.4}Evaporation residue}

In the DNS approach, the partial cross sections of the CN formation are used to calculate evaporation residue (ER) cross sections at given values of the CN 
excitation energy $E^*_{x}$ and angular momentum $\ell$:
\begin{eqnarray}\label{erpar}
    \sigma^x_{\rm ER}(E^*_{x})=\sum^{\ell_d}_{0}(2\ell+1)\sigma^x_{\rm ER}(E^*_{x},\ell).
\end{eqnarray}
where, $\sigma^x_{\rm ER}(E^*_{x},\ell)$ is the partial cross section of ER formation obtained after the emission of particles from the intermediate nucleus 
with the excitation energy $E^*_{x}$ at each step $x$ of the de-excitation cascade by the formula \cite{Nasirov2005, Mandaglio2012}:
\begin{eqnarray}\label{ercs}
    \sigma^x_{\rm ER}(E^*_{x},\ell)=\sigma^{x-1}_{\rm ER}(E^*_{x-1},\ell)W^x_{sur}(E^*_{x-1},\ell).
\end{eqnarray}
Here, $\sigma^{x-1}_{\rm ER}(E^*_{x-1},\ell)$ is the partial cross section of the intermediate excited nucleus formation at the $(x-1)$th step and obviously,  
$\sigma^{(0)}_{\rm ER}(E^*_{x-1},\ell)=\sigma_{fus}(E^*_{CN},\ell)$ which is calculated by (\ref{parfusav}); $W^x_{sur}(E^*_{x-1},\ell)$ is the survival 	
probability of the $x$th intermediate nucleus against fission along the de-excitation cascade of CN. The survival probability $W^x_{sur}(E^*_{x-1},\ell)$ is 
calculated by the statistical model implanted in KEWPIE2 \cite{Kewpie2}, which is dedicated to the study of SHEs. In presented results, the Weisskopf 
approximation \cite{Weisskopf} is used to calculate neutron emission width:
\begin{eqnarray}\label{WidthN}
    \Gamma_n=\frac{(2S_n+1)\mu_n}{\pi^2\hbar^2}\int\limits_0^{E^*_{CN}-B_n}\frac{\sigma^n_{inv}(\epsilon_n)\rho_B(E^*_B)\epsilon_nd\epsilon_n}{\rho_{CN}(E^*_{CN})}\nonumber\\
\end{eqnarray}
where $\rho_{CN}(E^*_{CN})$ is the level density of intermediate nucleus, $E^*_B$ is the excitation energy of the residual nucleus after the emission of a 
neutron, $B_n$ is the binding energy of the neutron with the residue nucleus, $S_n$ and 	$\epsilon_n$ denote the intrinsic spin of the neutron and its kinetic 
energy, $\sigma^n_{inv}$ is the cross section for the time-reversed reaction.
	
In the KEWPIE2 code, the fission-decay width is estimated within the standard Bohr–Wheeler transition-state model \cite{Wheeler}:
\begin{eqnarray}\label{WidthFis}
\Gamma_f^{BW}&=&\frac{1}{2\pi\rho_{CN}^{gs}(E^*_{CN},J_{CN})}\int\limits^{E^*_{CN}-B_f}_{0}\rho_C^{sd}(E^*_{sd},J_{CN})\nonumber\\ 
&&\times T_{fiss}(\epsilon_f)d\epsilon_f,
\end{eqnarray}
here, the excitation energy at the saddle point is equal $E^*_{sd}=E^*_{CN}-B_f-\epsilon_f$, $\rho_{CN}^{gs}(E^*_{CN},J_{CN})$ and $\rho_C^{sd}(E^*_{sd},J_{CN})$
 are the level densities of the nucleus at the ground-state and saddle-point deformation. In the KEWPIE2 code \cite{Kewpie2}, the improved state-density 
formula, which was first proposed in Ref. \cite{Grossjean1985}, has been employed to estimate various decay widths. $T_{fiss}(\epsilon_f)$ is a penetration factor which corresponds to the Hill-Wheeler transmission coefficient \cite{Wheeler1953}. The fission barrier consists of a 
macroscopic liquid-drop component and microscopic shell correction energy component. For SHEs, the macroscopic part is relatively small, and the liquid-drop 
model gives the fission barriers of less than 1 MeV for $Z>105$. The main contribution is given by the shell correction energy. However, the shell correction 
energies for SHEs have not yet been firmly confirmed \cite{zhang2013}. In KEWPIE2, a shell correction factor $f$ is introduced to account for this uncertainty \cite{Kewpie2}:
\begin{equation}
    B_{f}=B_{LD}-f\delta{W}
\end{equation}
where $B_{LD}$ and $\delta{W}$ are the liquid-drop fission barrier and the effective shell-correction energy, respectively. The liquid-drop fission barrier is 
estimated by using Lublin-Strasbourg Drop model \cite{Lubin}. The ground-state shell correction energies and the parameterizations for liquid-drop fission 
barrier are using the mass table of M\"{o}ller et al. \cite{Moller1995}. The dependence of the  fission barrier on the excitation energy $E^*_{CN}$ and 
angular momentum $\ell$ of the CN can be taken into account as in Ref.~\cite{Giardina2018}, where the correction factor $f$ was written as $h(T) q(\ell)$,  
\begin{eqnarray}\label{hT}
    h(T)=\{1+\exp[(T-T_0)/d]\}^{-1}
\end{eqnarray}
and
\begin{eqnarray}\label{qL}
    q(\ell)=\{1+\exp[(\ell-\ell_{1/2})/\Delta\ell]\}^{-1}.
\end{eqnarray}
In Eq.(\ref{hT}) $T=\sqrt{E^*_{CN}/a}$ is nuclear temperature, $d=0.3$ MeV is the rate of washing out the shell corrections with the temperature, 
$T_0=1.16$ MeV is the value at which the damping factor $h(T)$ is reduced by $1/2$; in Eq.(\ref{qL}), $\Delta\ell=3\hbar$ is the rate of washing out the 
shell corrections with the angular momentum, $\ell_{1/2}=20\hbar$ is the value at which the damping factor $q(\ell)$ is reduced by $1/2$. To calculate the 
level density parameter $a$, Ignatyuk’s prescription \cite{Ignatyuk1975} was used. 

\section{\label{Discussion}Results and discussion}

The evaporation residue cross sections for the $^{48}$Ca+$^{243}$Am and $^{54}$Cr+$^{243}$Am reactions have been calculated by the use of partial 
fusion cross sections found by Eq. (\ref{parfusav}). This way allows us to take into account the influence of the orbital angular momentum on the probability 
of the compound nucleus formation. It should be noted that the averaging procedures  over orientation angles ($0^o, 15^o, 30^o, 45^o, 60^o, 75^o$, and 
$90^o$) of the axial symmetry axis of the deformed  nuclei  $^{54}$Cr and $^{243}$Am by Eq. \ref{parfusal} and over 7 vibrational states ($-\beta^+_2,-2\beta^+_2/3,
-\beta^+_2/3,0,\beta^+_2/3,2\beta^+_2/3,\beta^+_2$) of the spherical 
nucleus $^{48}$Ca Eq. (\ref{parfusav}) have been performed to get the partial cross sections under discussion. 
  
\begin{figure}[htb]
\hspace*{-0.7cm}
\includegraphics[width=0.63\textwidth]{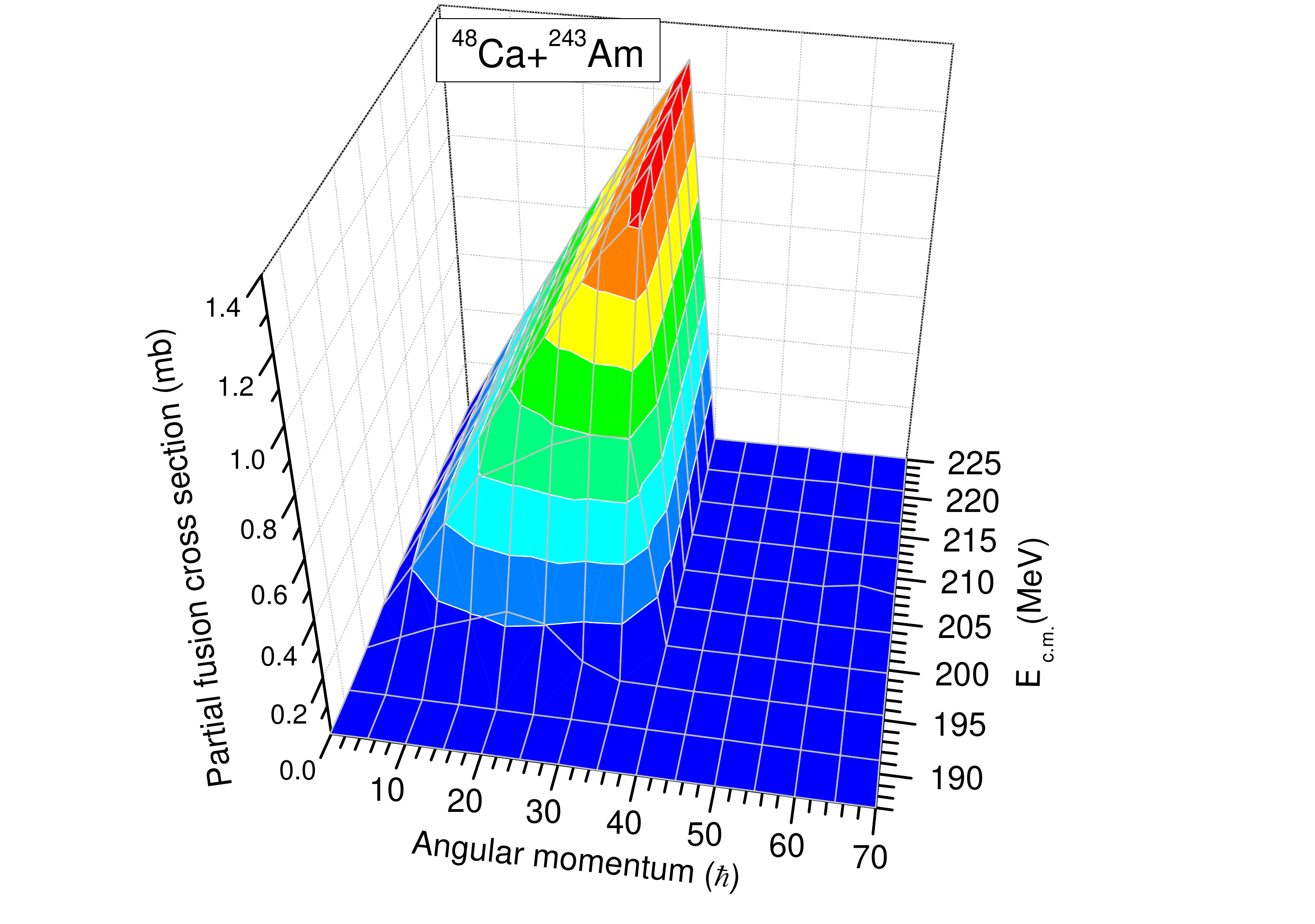}
\caption{(Color online) The  partial fusion cross section calculated in this 
work for the 
$^{48}$Ca+$^{243}$Am reaction as a function of the collision energy $E_{\rm c.m.}$ 
and orbital angular momentum $L$.} 
\label{PCFCS}
\end{figure}

\begin{figure}[htb]
\hspace*{-1.2 cm}
\includegraphics[width=0.58\textwidth]{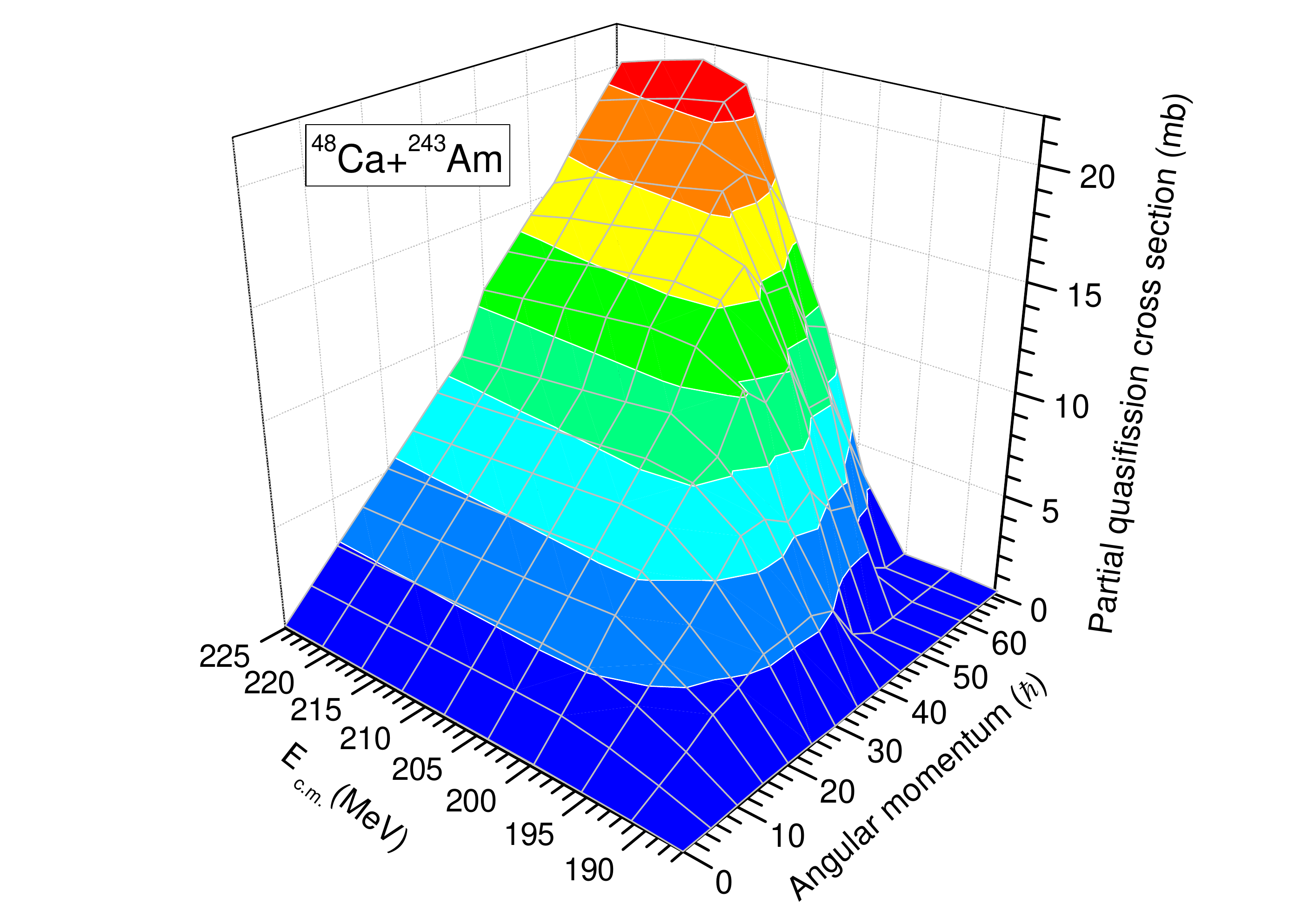}
\caption{(Color online) The same as in Fig. \ref{PCFCS} but for the quasifission
cross section.}
\label{PQFCS} 
\end{figure}

The results of the partial fusion and quasifission cross sections obtained for the $^{48}$Ca+$^{243}$Am reaction are presented in Fig. \ref{PCFCS} and 
\ref{PQFCS}, respectively. Two main differences are seen from these Figs: 1) the values of the quasifission cross section are more than one order larger 
the ones of complete fusion; 2) very wide range of the angular momentum values contribute to the formation of the quasifission products while the only values 
angular momentum up to $\ell=40$ can contribute the complete fusion cross sections since the fission barrier $B_f$ disappears at $\ell>40$. The part of the 
partial cross section corresponding to complete fusion and having angular momentum distribution $\ell>40$ can be considered as the cross section presenting 
the fast fission process. It should be stressed the set of DNS with the angular momentum $\ell>40$  has survived against quasifission and such DNS transform 
into mononucleus but it cannot reach compound nucleus state since at this condition it has not fission barrier $B_f$ providing possibility to have an equilibrated state.
	
\begin{figure}[htb]
\hspace*{-0.3cm}
\includegraphics[width=0.56\textwidth]{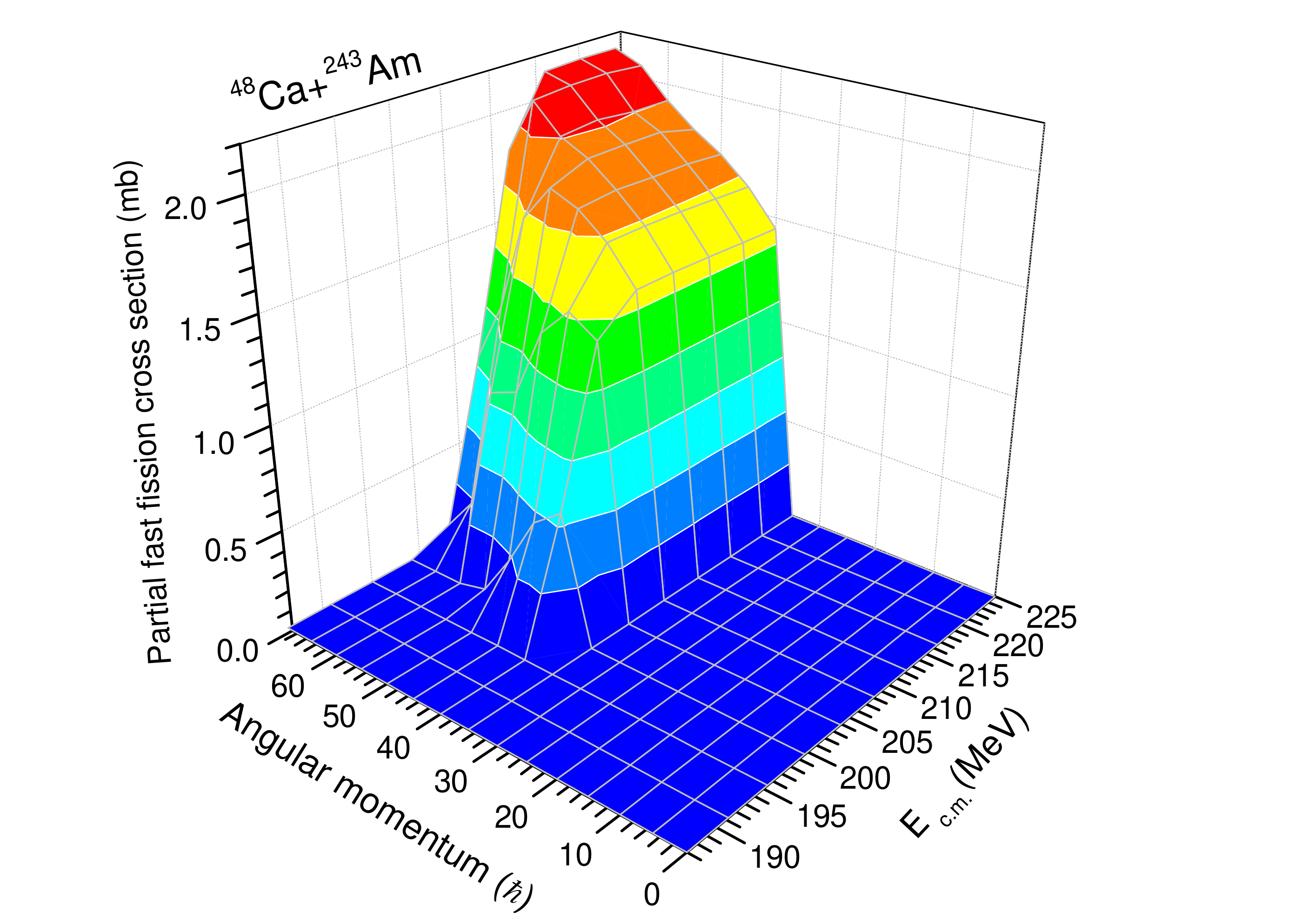}
\caption{(Color online) The same as in Fig. \ref{PCFCS} but for the  fast fission
cross section.}
\label{PFFCS}
\end{figure}

\begin{figure}[htb]
\hspace*{-0.5cm}
\includegraphics[width=0.54\textwidth]{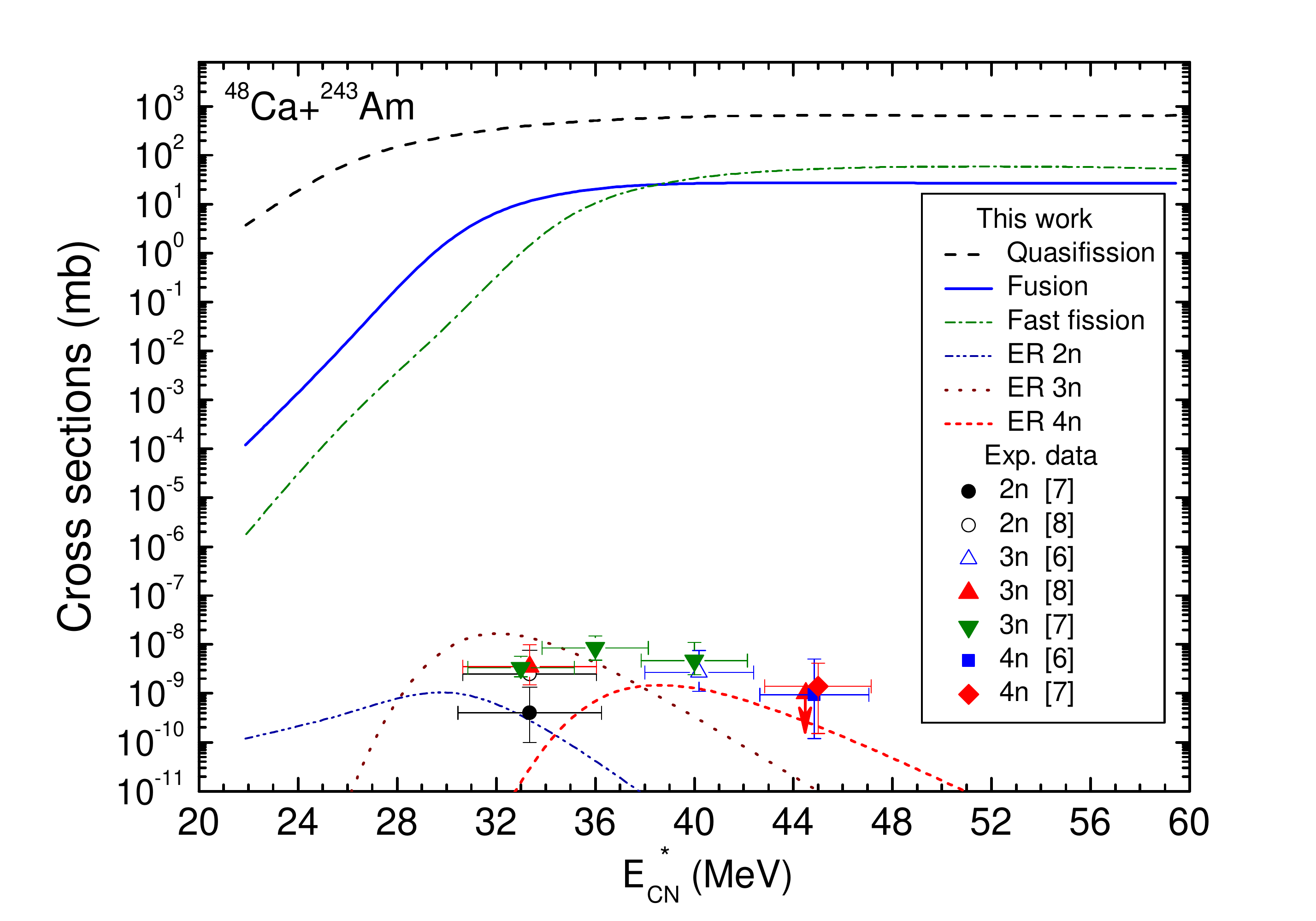}
\caption{(Color online) Theoretical cross sections of the quasifission 
(dashed black curve),
complete fusion (solid  blue curve), fast fission (dot-dashed green curve) 
and ER formation in 2n (thin dot-dashed violet curve), 3n (thin wine dotted curve)
 and 4n (thin short dashed red curve) channels, 
as well as the experimental data of the ER formation in the same channels 
as a function of the CN excitation energy: the 2n channel data labelled by solid 
and open circles have been obtained from Refs.~\cite{Oganessian2012} and
 ~\cite{Oganessian2013}, respectively; the 3n channel data shown by 
 the open up and down triangles were obtained from Refs. ~\cite{Oganessian2005}
 and \cite{Oganessian2013}; the 4n channel data shown by the square and diamond 
 were obtained from Refs.~\cite{Oganessian2005} and ~\cite{Oganessian2012}, 
 respectively.}
\label{CaAmCS}      
\end{figure}

The angular momentum distribution of the mononucleus which is a set of the DNS survived against quasifission is demonstrated in Fig. \ref{PFFCS}. The 
fission of mononucleus with the angular momentum $\ell>\ell_f$ is called fast fission. For the $^{48}$Ca+$^{243}$Am reaction it occurs at $\ell \geq 40$. 
The damping the fission barrier is calculated by Eqs.(\ref{qL}) and (\ref{hT}). 
At last the cross sections for the yield of the evaporation residues after emission of 2, 3 and 4 neutrons (2n, 3n and 4n channels) in competition against fission 
have been calculated for the wide range of the excitation energy of the compound nucleus. The theoretical results of this work is compared with the 
experimental data obtained in Refs.~\cite{Oganessian2005,Oganessian2012,Oganessian2013}. The comparison of the experimental data  and  theoretical 
results of the ER cross sections, as well as the theoretical results obtained in this work for the complete fusion, quasifission and fast fission processes are 
presented in Fig.~\ref{CaAmCS}. It is seen that the agreement between theoretical and experimental results for the 2n, 3n and 4n channels of the ER 
formation are good. The values of the $E^*_{\rm CN}$ are estimated only theoretically by the use of the reaction energy balance $Q_{gg}$ and collision 
energy $E_{\rm c.m.}$:  $E^*_{\rm CN}=E_{\rm c.m.}+Q_{gg}$, where $Q_{gg}=B_P(^{48}$Ca)+$B_T(^{243}$Am)-$B_{CN}(^{291}$Mc) depends on 
the binding energies ($B_P$ and $B_T$) of the colliding nuclei and being formed compound nucleus ($B_{CN}(^{291}$Mc)). The value of $B_{CN}(^{291}$Mc)is equal to 178.97 and 182.99 MeV in Refs. \cite{Moller1995} and \cite{Muntian2003}, respectively. This means that the value of $E^*_{\rm CN}$ 
corresponding  to the given value $E_{\rm c.m.}$ may be different in dependence on what the mass table is used by the authors of papers. In 
Ref.~\cite{Oganessian2013} the spreading $E^*_{\rm CN}$ is presented in their Table 1. We have presented the ranges for the $E^*_{\rm CN}$  values at 
the given beam energies $E^*_{\rm lab}$ in Table \ref{tabEcn}. As it is seen from the Table \ref{tabEcn} and Fig. \ref{CaAmCS} there is an uncertainty in 
estimation of the CN excitation energy with the width about 4 MeV. The absolute values of the theoretical and experimental data are in a good agreement.
\begin{table}[b]
\caption{\label{tabEcn}%
Laboratory-frame beam energies $E_{\rm lab}$ in the middle of the target layers, in the mass-centre system collision energies $E_{\rm c.m.}$ resulting excitation energy intervals,  $E^*_{\rm CN}$ values calculated by the use of the mass tables in~\cite{Moller1995} and~\cite{Muntian2003}, respectively. }
\begin{ruledtabular}
\begin{tabular}{ccccc}
$E_{\rm lab}$ & $E_{\rm c.m.}$ & CN excitation energy & $E^*_{\rm CN}$~\cite{Moller1995} & $E^*_{\rm CN}$~\cite{Muntian2003}\\
        (MeV) &   (MeV)      & intervals (MeV)~\cite{Oganessian2012,Oganessian2013} & (MeV) &   (MeV)\\
\colrule
239.8 & 200.25 & 31.1–-35.3   & 34.23 & 30.21\\
240.5 & 200.83 & 31.4–-36.2   & 34.81 & 30.79\\
241.0 & 201.25 & 32.0–-36.4   & 35.23 & 31.21\\
243.4 & 203.25 & 34.0–-38.3   & 37.23 & 33.21 \\
248.1 & 207.18 & 38.0–-42.3   & 41.16 & 37.14 \\
253.8 & 211.94 & 42.5–-47.2   & 45.92 & 41.90 \\
\end{tabular}
\end{ruledtabular}
\end{table}
\begin{figure}[htb]
\hspace*{-0.3cm}
\includegraphics[width=0.48\textwidth]{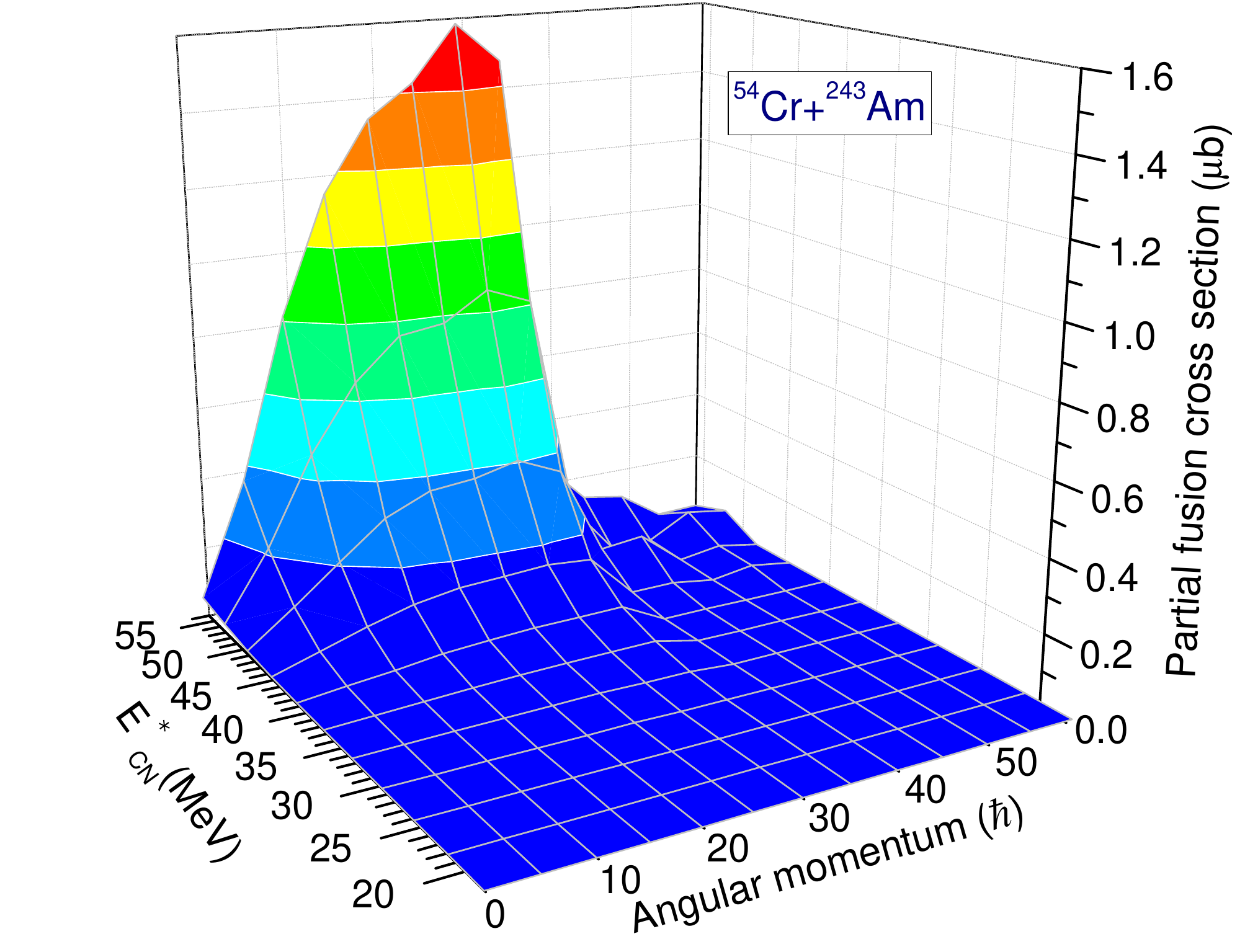}
\caption{(Color online) The  partial fusion cross section calculated in this 
work for the  $^{54}$Ca+$^{243}$Am reaction as a function of the collision energy $E_{\rm c.m.}$ and orbital angular momentum $L$.}
\label{PCFCS54Cr243Am}
\end{figure}
\begin{figure}[htb]
\hspace*{-0.3cm}
\includegraphics[width=0.47\textwidth]{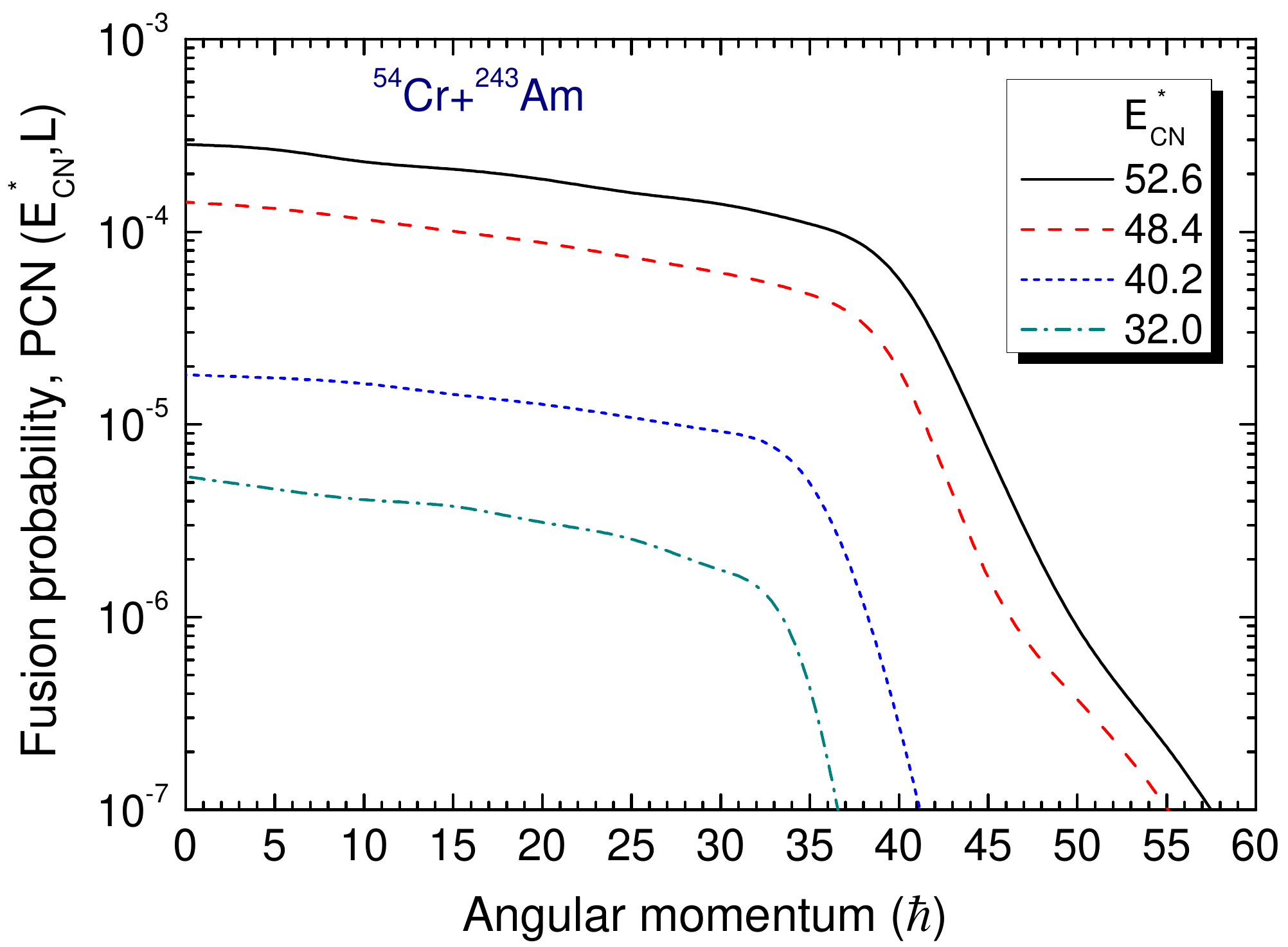}
\caption{(Color online) 
The dependence of the fusion probability $P_{\rm CN}(E_{\rm c.m.}, L)$ on the 
orbital angular momentum for the values of the CN excitation energies  
$E^*_{\rm CN}$=32.0, 40.2, 48.4 and 52.6 MeV for 
the $^{54}$Cr+$^{243}$Am reaction.}
\label{PcnEL}
\end{figure}
\begin{figure}[htb]
\centering
\includegraphics[width=0.48\textwidth]{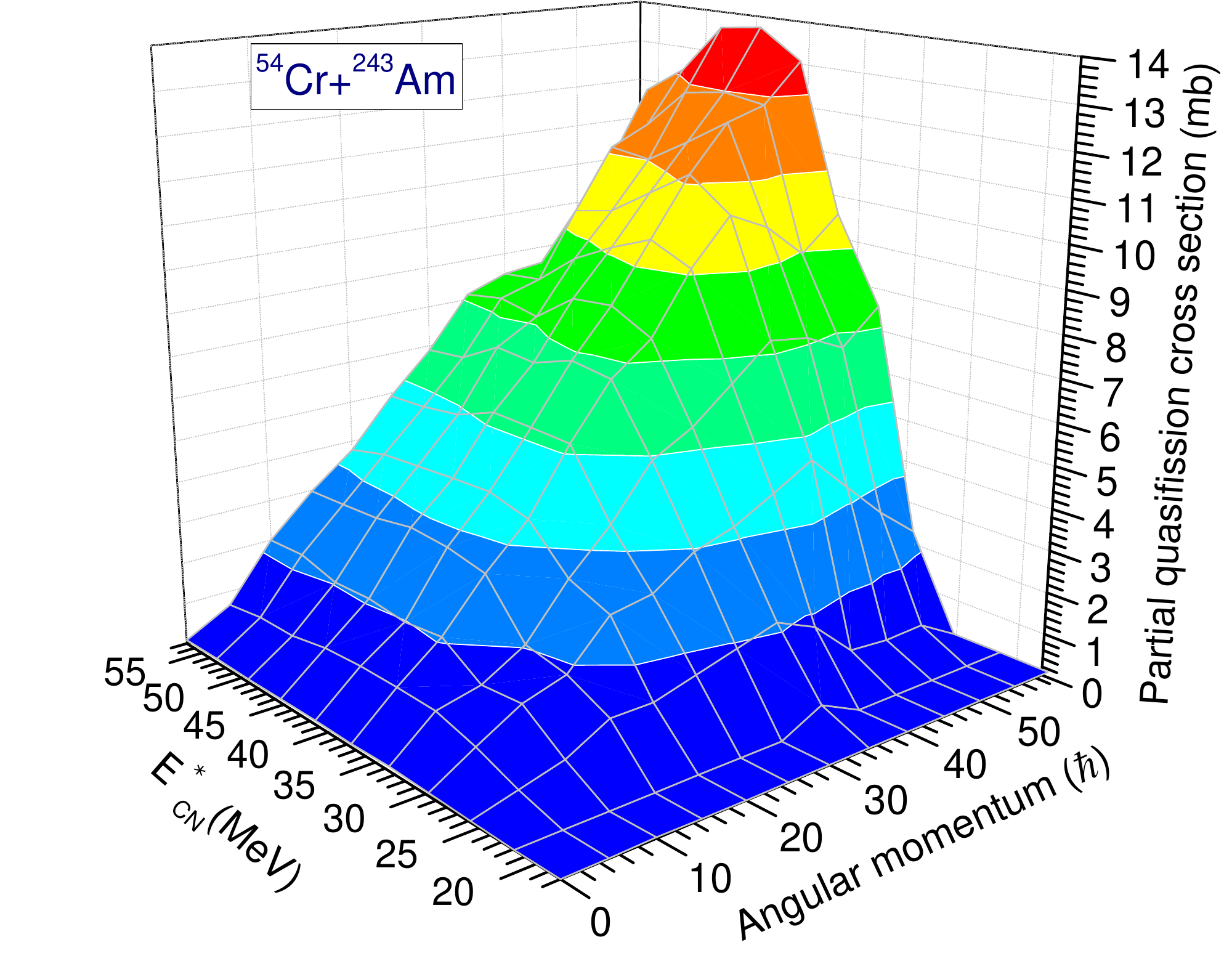}
\caption{(Color online) 
 The same as in Fig. \ref{PCFCS54Cr243Am} but for the quasifission
cross section.}
\label{PQFCS54Cr243Am}
\end{figure}
\begin{figure}[htb]
\hspace*{-0.3cm}
\includegraphics[width=0.47\textwidth]{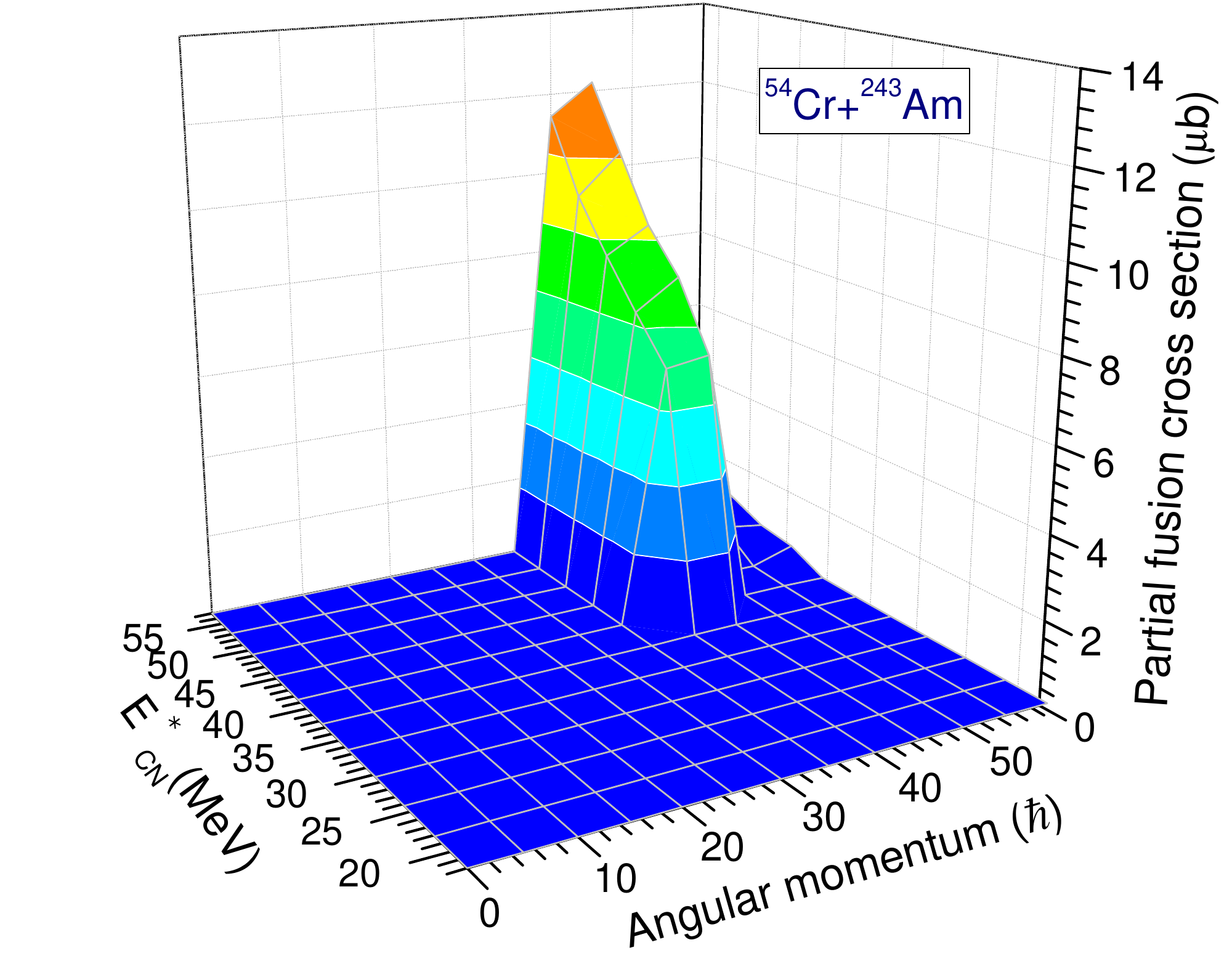}
\caption{(Color online) 
 The same as in Fig. \ref{PCFCS54Cr243Am} but for the fast fission
cross section.}

\label{PFFCS54Cr243Am}
\end{figure}
\begin{figure}[htb]
\centering
\includegraphics[width=0.47\textwidth]{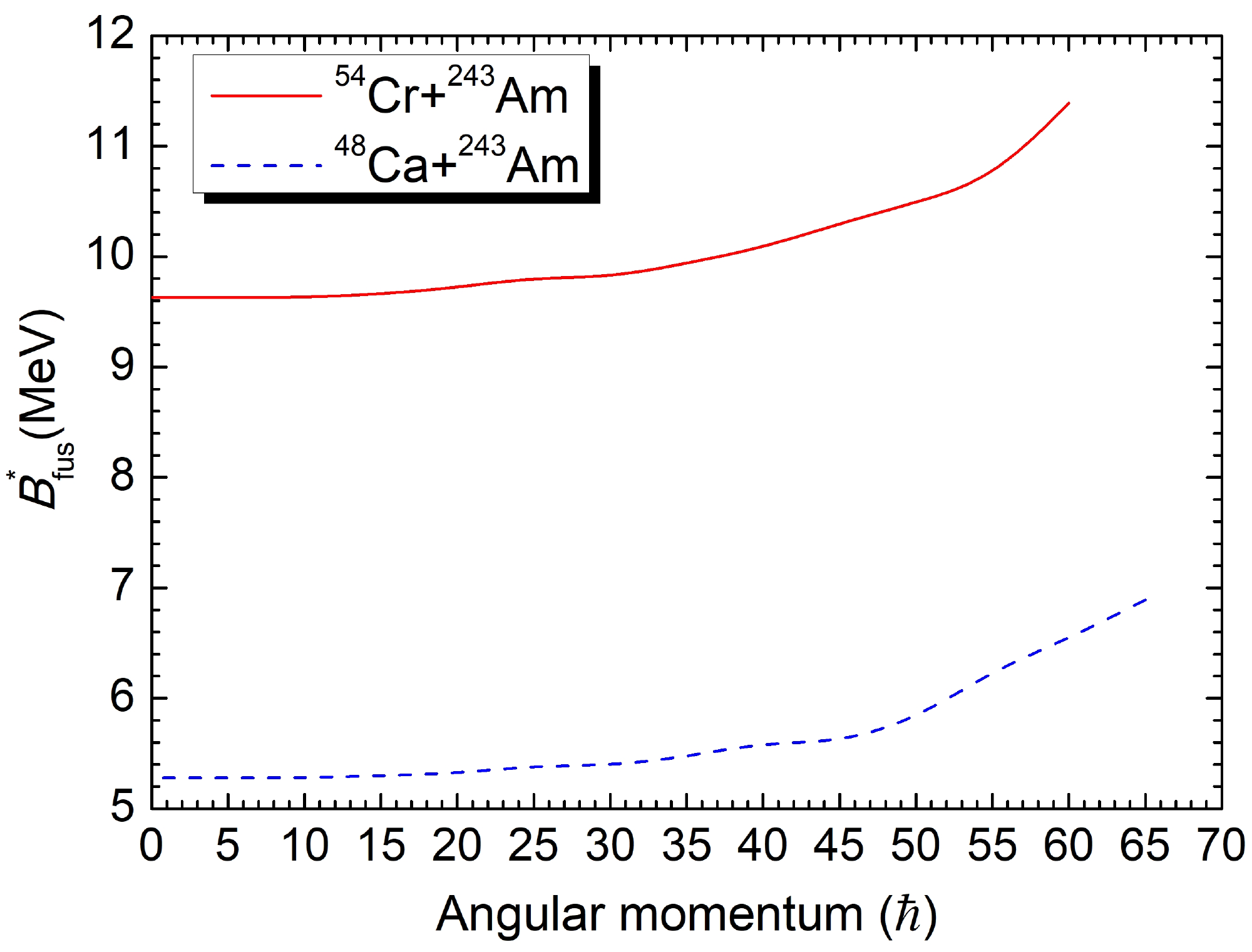}
\caption{(Color online) Comparison of the results for the intrinsic fusion 
barrier $B^*_{\rm fus}$ for the $^{54}$Cr+$^{243}$Am and $^{48}$Ca+$^{243}$Am 
 reactions calculated in this work as a function of the DNS angular momentum 
 ($L$).}
\label{BfusL}
\end{figure}
\begin{figure}[htb]
\hspace*{-0.3cm}
\includegraphics[width=0.47\textwidth]{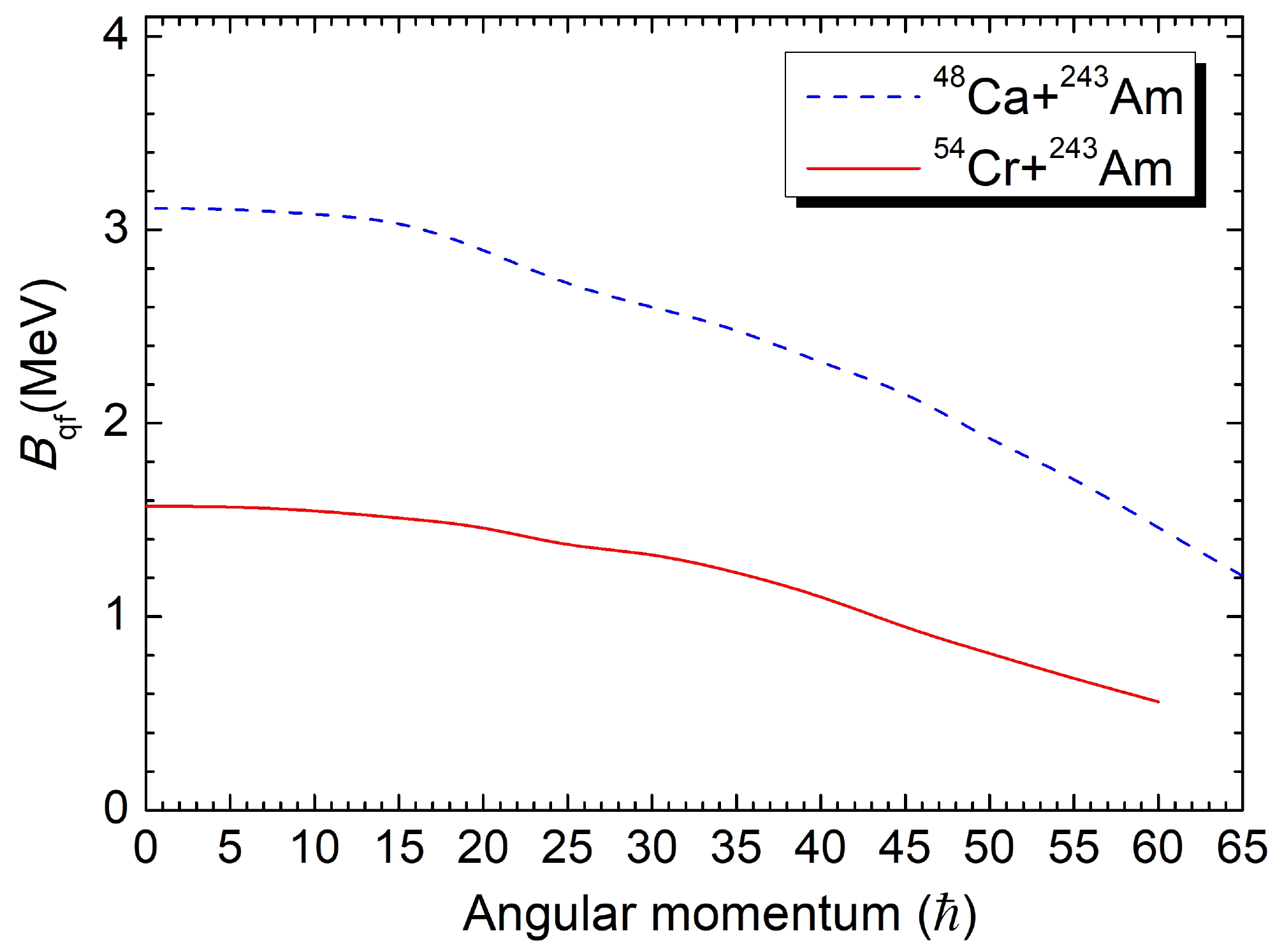}
\caption{(Color online) Comparison of the results for the quasifission barrier
$B_{qf}$ for the $^{54}$Cr+$^{243}$Am and $^{48}$Ca+$^{243}$Am  reactions 
calculated in this work as a function of the DNS angular momentum ($L$).}
\label{BqfL}
\end{figure}
\begin{figure}[htb]
\hspace*{-0.3cm}
\includegraphics[width=0.47\textwidth]{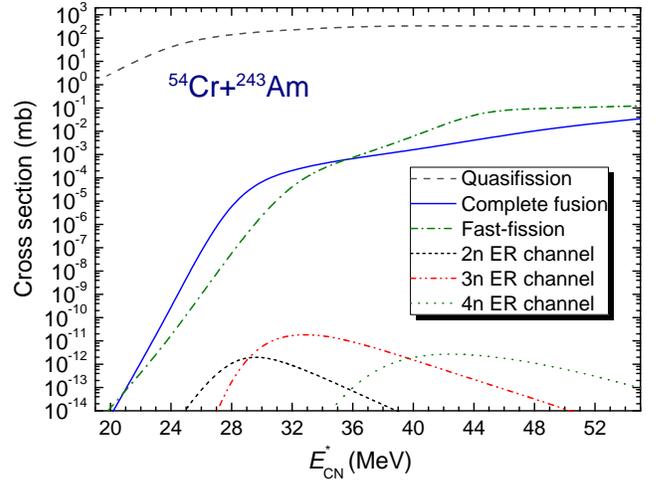}
\caption{(Color online) 
Theoretical cross sections of the quasifission (dashed black curve),
complete fusion (solid  blue curve), fast fission (dot-dashed green curve) 
and ER formation in 2n (thin short dashed black curve), 
3n (thin dash-double dotted red curve)
 and 4n (thin  dotted green curve) channels cross sections 
 calculated in this work as a function of $E^*_{CN}$ 
 for the $^{54}$Cr+$^{243}$Am reaction.}
\label{CS}
\end{figure}
\begin{figure}[htb]
\hspace*{-0.3cm}
\includegraphics[width=0.48\textwidth]{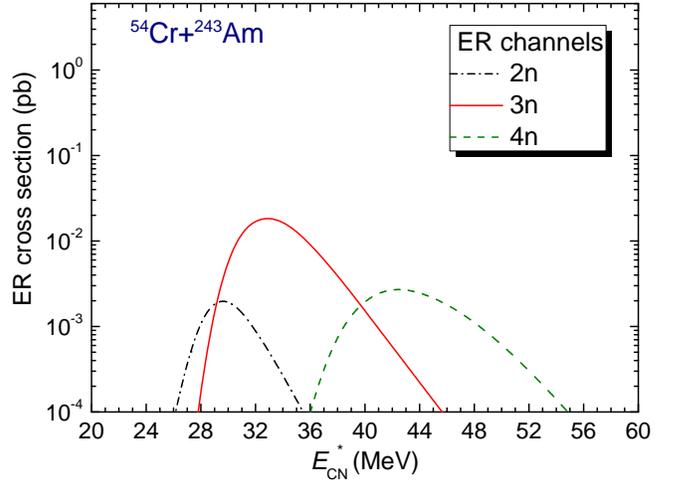}
\caption{(Color online) The expected ER cross section for 
producing superheavy element 119 via $^{54}$Cr+$^{243}$Am reaction 
in the 2n (dot-dashed black curve), 3n (solid red curve) and 
4n (dashed green curve) channels.}
\label{ER54Cr243Am} 
\end{figure}
The successful application of the method presented in Section 3 to the description of the ER cross sections of the $^{48}$Ca+$^{243}$Am allows us to calculate the ER cross sections of the $^{54}$Cr+$^{243}$Am which leads to formation  of the SHE
 with the charge number $^{297}$119.
	
In Fig. \ref{PCFCS54Cr243Am}, the partial fusion cross section (in units $\mu$b) calculated by Eq. (\ref{parfusav}) for the  $^{54}$Cr+$^{243}$Am reaction 
is presented as a function of the CN excitation energy $E^*_{CN}$ and orbital angular momentum $\ell$. The sharp decreasing its values at $\ell > 35$ is 
related with the disappear of the fission barrier of the compound nucleus $^{297}$119. This phenomenon is taken into account by Eqs. (\ref{hT}) and (\ref{qL}) 	which give a decrease of the fission barrier $B_f$ by increase of the excitation energy and angular momentum of the compound nucleus formed at 
complete fusion of colliding nuclei.  The mononucleus (the DNS being transformed into compound nucleus)  with the angular momentum distributions in the 
range larger than $\ell > 35$  undergoes to fast fission 	and its products can have mass distributions as ones of fusion-fission and/or quasifission. The 
dependence of the fusion probability $P_{\rm CN}$ on the orbital angular momentum for the values of the CN excitation energies  $E^*_{\rm CN}$=32.0,   
40.2, 48.4 and 52.6 MeV is shown in Fig. \ref{PcnEL}. The increase of the $P_{\rm CN}$ values by the increase of the collision energy is due to the increase 
of DNS excitation  energy and its decreasing by the increase of the angular momentum is caused by the increase of the intrinsic fusion barrier $B^*_{fus}$ as 
a function of the angular momentum. The decrease of the fission barrier leads to the  strong decrease of $P_{\rm CN}$ since the being formed CN becomes 
less stable against fission. Very small values of the fusion cross section obtained for the $^{54}$Cr+$^{243}$Am in comparison with the ones calculated for 
the $^{48}$Ca+$^{243}$Am reactions are due to large values of the intrinsic fusion barrier $B^*_{\rm fus}$ for the former reaction, as well as the decrease 
of the quasifission barrier due to increase the charge number  $Z=24$ of $^{54}$Cr  against the charge number $Z=20$ of $^{48}$Ca in the entrance 
channel. 

The partial cross section of the quasifission process (see Fig. \ref{PQFCS54Cr243Am})  is much larger in comparison with the cross sections of the fusion and fast fission products since according to the results presented in Figs. \ref{PCFCS54Cr243Am} and  \ref{PFFCS54Cr243Am}. This means the  yield 
of the quasifission products is dominant among the yield of the products of the other mechanisms. Here we should note that the unit of the quasifission cross 
section is in millibarn (mb) in Fig. \ref{PQFCS54Cr243Am} while  the units of the fusion (Fig. \ref{PCFCS54Cr243Am}) and fast fission cross sections (Fig. \ref{PFFCS54Cr243Am}) are 
presented in microbarn  $\mu$b. 
To answer on the question about reasons 
calling the difference  between the cross sections of the complete fusion of the $^{48}$Ca+$^{243}$Am and   $^{54}$Cr+$^{243}$Am reactions we have  
presented in Figs. \ref{BfusL} and \ref{BqfL} the comparisons of the intrinsic  fusion $B^*_{\rm fus}$ and quasifission $B_{\rm qf}$  barriers, respectively, 
calculated in this work. It is seen from Fig.~\ref{BfusL} that the hindrance to complete fusion is stronger in the  $^{54}$Cr+$^{243}$Am reaction in comparison    
with the case of the $^{48}$Ca+$^{243}$Am reaction since the values of $B^*_{\rm fus}$ for the former reaction are about two times larger the ones for 
the latter reaction. At the same time the values of the $B_{\rm qf}$ for the  $^{54}$Cr+$^{243}$Am reaction are smaller about two times in comparison  
with the case of the $^{48}$Ca+$^{243}$Am reaction. Certainly, the DNS lifetime formed in the  $^{48}$Ca+$^{243}$Am reaction is longer it has more 
opportunity to be transformed into compound nucleus. 	   

The total cross sections for the  $^{54}$Cr+$^{243}$Am reaction summarized over values of the  angular momentum are presented in Fig. \ref{CS}. It is 
seen that the cross sections of the complete fusion and fast fission are comparable.

The excitation functions  of the 2n, 3n and 4n de-excitation channels are presented in Fig. \ref{CS} and separately in Fig. \ref{ER54Cr243Am}. The maximum 
of the evaporation residue cross section predicted around the excitation energy of compound nucleus $E^*_{\rm CN}$= 32--35 MeV and 3n channel of the 
evaporation residues events dominates. Maximum value of its cross section is about 25 fb and the corresponding events 
can be observed in the 
SuperHeavy Element Factory (DC-280 cyclotron) of the Flerov Laboratory of Nuclear Reaction (JINR, Dubna) \cite{Shefactory}. 
In Ref.~\cite{Li2018}, the predicted cross section for the 3n channel of ER for the $^{53}$Cr+$^{243}$Am  reaction is 0.95 pb at the excitation energy $E^*_{\rm CN}$=33 MeV. Looking at these values  we can conclude that our prediction is intermediate between them. Authors of 
Ref.~\cite{XingLv2021} have predicted the maximum value of the ER cross section of the 3n channel for the $^{54}$Cr+$^{243}$Am reaction for 1.82 fb.   

\section{Conclusion}

The peculiarities of the processes in the heavy ion reactions, which can lead to formation of superheavy elements, have been explored by the methods of the 
DNS  concept. The formation probability of the DNS at capture stage of the heavy ion collisions is calculated by solution of the equation of motions for the 
relative distance and orbital angular momentum at the given collision energy $E_{\rm c.m.}$ for the  $^{48}$Ca+$^{243}$Am and $^{54}$Cr+$^{243}$Am 
reactions. An orientation angle of the symmetry axis of the deformed nucleus and vibrational states of the spherical nucleus are included in calculation of the 
partial capture cross sections.   

The transformation probability of the DNS into compound nucleus is calculated by the estimations of the branching ratio of the nuclear level densities on the 
saddle point to the complete fusion  and on the Coulomb barrier of the nucleus-nucleus interaction of the potential energy surface. The results of calculation 
have shown the quasifission is dominant channel of the capture events for the both reactions. It has very large cross sections for the  $^{54}$Cr+$^{243}$Am reaction. The probability of the observation of the new superheavy element $Z$=119 enough small and it is near border of the experimental possibilities.     
Our calculations have given the result about 0.025 pb for the evaporation residue cross section at the excitation energy $E^*_{\rm CN}$=32 MeV for the 3n-
channel. To show the dependence of the all stages of the reaction chain at formation of superheavy element on the DNS angular momentum the partial cross 
sections  of the quasifission, complete fusion and fast fission processes are presented for the wide range of the excitation energy of the compound nucleus. 
The reasons leading  to the dominance of the quasifission over complete fusion and fast fission processes and causing strong difference in the cross sections 
of complete fusion for the  $^{48}$Ca+$^{243}$Am and $^{54}$Cr+$^{243}$Am reactions are the values of the intrinsic fusion barrier calculated for the 
wide range of the DNS angular momentum. It is demonstrated that  the values of the intrinsic fusion barrier for the  $^{54}$Cr+$^{243}$Am reaction is 
about two times higher than ones obtained for the $^{48}$Ca+$^{243}$Am reaction. Therefore, a hindrance to complete fusion is stronger for the 
reactions with the more massive  projectile $^{54}$Cr. The DNS formed in the $^{54}$Cr+$^{243}$Am reaction is less stable in comparison with the one 
formed in the $^{48}$Ca+$^{243}$Am reaction since the quasifission barriers for the DNS   calculated for the former reaction is smaller than the ones found 
for the latter reaction.

\begin{acknowledgments}
The authors are grateful to the Ministry of Innovative Development of Uzbekistan (M.I.D.U.), for financial support in the implementation of this research work.
\end{acknowledgments}

\bibliographystyle{apsrev4-2}
\bibliography{References_PRC}

\end{document}